CHAOKANG TAI*

# Left Radicalism and the Milky Way: Connecting the Scientific and Socialist Virtues of Anton Pannekoek

## ABSTRACT

Anton Pannekoek (1873–60) was both an influential Marxist and an innovative astronomer. This paper will analyze the various innovative methods that he developed to represent the visual aspect of the Milky Way and the statistical distribution of stars in the galaxy through a framework of epistemic virtues. Doing so will not only emphasize the unique aspects of his astronomical research, but also reveal its connections to his left radical brand of Marxism. A crucial feature of Pannekoek's astronomical method was the active role ascribed to astronomers. They were expected to use their intuitive ability to organize data according to the appearance of the Milky Way, even as they had to avoid the influence of personal experience and theoretical presuppositions about the shape of the system. With this method, Pannekoek produced results that went against the Kapteyn Universe and instead made him the first astronomer in the Netherlands to find supporting evidence for Harlow Shapley's extended galaxy. After exploring Pannekoek's Marxist philosophy, it is argued that both his astronomical method and his interpretation of historical materialism can be seen as strategies developed to make optical use of his particular conception of the human mind.

*Institute for Theoretical Physics Amsterdam, Anton Pannekoek Institute for Astronomy, and Vossius Center for History of Humanities and Sciences, University of Amsterdam, Science Park 904, PO Box 94485, 1090 GL Amsterdam, the Netherlands; Descartes Centre for the History and Philosophy of the Sciences and the Humanities, Utrecht University, the Netherlands; cktai@uva.nl

The following abbreviations are used: API, Archive of the Anton Pannekoek Institute for Astronomy, University of Amsterdam, Amsterdam; *ApJ*, *Astrophysical Journal*; *BAN*, *Bulletin of the Astronomical Institutes of the Netherlands*; *HSPBS*, *Historical Studies of the Physical and Biological Sciences*; *JHA*, *Journal for the History of Astronomy*; *Legacy*, Piet C. van der Kruit and Klaas van Berkel (eds.), *The Legacy of J.C. Kapteyn: Studies on Kapteyn and the Development of Modern Astronomy* (Dordrecht: Kluwer Academic, 2000); *MNRAS*, *Monthly Notices of the Royal Astronomical Society*; *PASP*, *Publications of the Astronomical Society of the Pacific*; *PKAW*, *Proceedings of the Section of Sciences*, Koninklijke Akademie van Wetenschappen te Amsterdam; *PUA*, *Publications of the Astronomical Institute of the University of Amsterdam*; *VS*, *Victorian Studies*; WdS, Leiden Observatory Archives, directorate Willem de Sitter, Leiden University Library, Leiden.

When Harvard University celebrated its tercentenary in 1936, it conveyed honorary degrees to 39 distinguished international scholars. Among the recipients was Dutch astronomer Anton Pannekoek (1873–1960), who was honored for his "contributions of high merit in many fields of astronomy."[1] Among other things, Pannekoek determined the distances to several Milky Way clouds, providing early evidence of the eccentric position of the Solar System in our galaxy; computed the first curve of growth for a star other than the Sun; and produced accurate drawings of the Milky Way, which the Carl Zeiss company in Jena would feature in their planetariums for decades.[2] There was also a very different side to Pannekoek, however. Prior to attending the tercentenary conference, he addressed the members of the Workers' Socialist Party of the United States, a small but active left communist movement in Boston.[3] Their members, who turned out in large numbers for the meeting, had a very different reason to be interested in Pannekoek. They knew him as former party theoretician of the Sozialdemokratische Partei Deutschlands (SPD), the largest socialist party of its time, and as one of the current theoretical leaders of the council communist movement, which they supported.

Throughout history, few people have managed to be as influential as Pannekoek in such widely different fields as Marxism and astronomy. The fact that he managed to contribute significantly to both fields makes him a compelling case study for the history of science. Previous research on the interaction between Marxism and science has often focused either on how Soviet scientists interacted with state ideology[4] or on how leftist scientists combined their socialist and scientific activities.[5] In the case of Pannekoek, however, it is possible to investigate the intellectual relation between scientific research and political ideology through the beliefs and practices of a single person. Recent scholarship has shown that such an approach can often provide new perspectives on specific methodological choices and epistemic beliefs.[6] This raises the question whether the same can be achieved in the case of Pannekoek. Can we get a better understanding of Pannekoek's astronomy if we investigate it in relation to his Marxism?

---

1. "The Harvard Tercentenary," *Science* 83 (1936): 566–67, on 567.

2. Carl Zeiss Jena to Anton Pannekoek, 25 Jul 1927, API; H. C. King, "The London Planetarium," *The Observatory* 78 (1958): 69–72, esp. 70.

3. Isaac Rabinowich to Anton Pannekoek, 5 Sep 1936, IISH inv. nr. 53/1; Klara Doris Rab, *Role-Modeling Socialist Behavior: The Life and Letters of Isaac Rab* (Raleigh, NC: Lulu.com, 2010), esp. 56 and 202.

4. See, e.g. David Joravsky, *Soviet Marxism and Natural Science* (New York: Columbia University Press, 1961); Kendall E. Bailes, "Science, Philosophy and Politics in Soviet History: The Case of Vladimir Vernadskii," *Russian Review* 40 (1981): 278–99; Loren R. Graham, *Science, Philosophy, and Human Behavior in the Soviet Union* (New York: Columbia University Press, 1987).

To explore this question, I will investigate the epistemic virtues that Pannekoek advocated in his astronomical writing. Epistemic virtues can be imagined as implicit moral guidelines for the practice of scientific research. They are epistemic because they indicate how knowledge should be extracted from observations and what role of the scientist plays this process. They are virtues because they prescribe the ideal behavior of the scientist—the scientific persona. Although epistemic virtues are often implicit, they are always normative, and as such, they are present beneath the surface of scientific publications: in the assessment of other researchers, in the explanation of method, or in the presentation of results. Likewise, traits ascribed to the scientific persona can be found by looking at how scientists praise or criticize their colleagues, and in the way that they themselves defend against criticism.

Recently, many scholars have taken up the task to investigate the epistemic virtues and the scientific or scholarly persona in historical studies of both science and the humanities.[7] One of the most prominent examples is Lorraine

---

5. See, e.g. Gary Werskey, *The Visible College: Scientists and Socialists in the 1930s* (London: Allen Lane, 1978).

6. See, e.g. Alexei Kojevnikov, "Freedom, Collectivism, and Quasiparticles: Social Metaphors in Quantum Physics," *HSPBS* 29 (1999): 295–31; Alexei Kojevnikov, "David Bohm and Collective Movement," *HSPBS* 33 (2002): 161–92; Anja Skaar Jacobsen, "Léon Rosenfeld's Marxist Defense of Complementarity," *HSPBS* 37, supplement (2007): 3–34; Peter Galison, "Assassin of Relativity," *Einstein for the 21st Century: His Legacy in Science, Art, and Modern Culture*, ed. Gerald Holton, Peter Galison, and Silvan S. Schweber (Princeton, NJ: Princeton University Press, 2008), 185–204.

7. See, e.g. Jessica Wang, "Physics, Emotion, and the Scientific Self: Merle Tuve's Cold War," *Historical Studies in the Natural Sciences* 42 (2012): 341–88; Gadi Algazi, "Exemplum and Wundertier: Three Concepts of the Scholarly Persona," *BMGN—Low Countries Historical Review* 131 (2016): 8–32; Herman Paul, "The Scholarly Self: Ideals of Intellectual Virtue in Nineteenth-Century Leiden," in *The Making of the Modern Humanities, Volume II: From Early Modern to Modern Disciplines*, ed. Thijs Weststeijn, Jaap Maat, and Rens Bod (Amsterdam: Amsterdam University Press, 2012), 397–411; Jeroen van Dongen, *Einstein's Unification* (Cambridge: Cambridge University Press, 2010).

Daston and Peter Galison's seminal book *Objectivity*, which traces the history of mechanical objectivity as an epistemic virtue. In particular, they state that mechanical objectivity was developed as a reaction to the reigning epistemic virtue: truth-to-nature, which placed emphasis on the natural philosopher as a genial sage who had to use reason and creativity to find the universal hidden behind the appearance. Steadily, however, the ideal vision of a scientist changed to one who operated in a machine-like manner to avoid the influence of his or her own subjectivity in an effort to let nature speak for itself. This was the virtue of mechanical objectivity. Mechanical objectivity itself was supplanted as the dominant virtue in the early twentieth century by the trained judgement, which called upon the educated professionals to use their trained intuition to search for structure and family resemblances in natural phenomena. Throughout *Objectivity*, Daston and Galison envision epistemic virtues as technologies of the self, developed to counteract the perceived weaknesses of the self and emphasize its strengths. Mechanical objectivity, for example, was tied to a conception of the self that was active and always imposed its subjectivity on observations. This active self could be tamed by using mechanical techniques of representation and self-restraint. Truth-to-nature on the other hand was connected with a fragmented self wherein the human mind was seen as a collection of faculties. To achieve truthfulness, it was imperative to call upon the faculties of reason and controlled imagination, while shutting out the passive imagination, which could lead to delusions and fanaticism.[8]

Daston and Galison's work was widely praised but not accepted without reservations. There were concerns regarding the validity of aligning the demise of mechanical objectivity with rising professionalism, the focus on scientific atlases rather than other images, and the large scope of their research. Most of all, however, reviewers have criticized the lack of social factors involved in the transformation of epistemic virtues.[9] In this paper, we will reflect on some of

8. Lorraine Daston and Peter Galison, *Objectivity* (New York: Zone Books, 2007). See also: Lorraine Daston and Peter Galison, "The Image of Objectivity," *Representations* 40 (1992): 81–128; Lorraine Daston, "Fear and Loathing in the Imagination of Science," *Daedalus* 127 (1998): 73–95; Peter Galison, "Image of Self," in *Things that Talk: Object Lessons from Art and Science*, ed. Lorraine Daston (New York: Zone Books, 2004); Lorraine Daston and Peter Galison, "Response: Objectivity and Its Critics," *VS* 50 (2008): 666–77.

9. This assessment is based on the following reviews: Theodore M. Porter, "The Objective Self," *VS* 50 (2008): 641–47; Jennifer Tucker, "Objectivity, Collective Sight, and Scientific Personae," *VS* 50 (2008): 648–57; John V. Pickstone "The Disunities of Representation," *British Journal for the History of Science* 42 (2009): 595–600; Nancy Anderson, "Eye and Image: Looking at a Visual Studies of Science," *Historical Studies in the Natural Sciences* 39 (2010): 115–25; Peter these issues through the case of Pannekoek. We will investigate if and how mesoscopic histories can help in understanding individual case studies, how Daston and Galison's mesoscopic story holds up when placed in comparison to a specific episode in the history of astronomy, and especially whether epistemic virtues can assist in understanding scientific research within the context of its social and cultural milieu.

Of course, one could wonder why we would want to use the framework of epistemic virtues for Pannekoek at all. There are good reasons, however, to believe that a focus on epistemic virtues can significant benefit individual case studies, especially when looking beyond the constraints of disciplinary boundaries. Matthew Stanley, for example, has shown how many of Arthur Eddington's epistemic virtues, like "seeking" and reliance on human experience, can be directly traced back to religious virtues he held as a Quaker.[10] In this paper, my approach will be slightly different. Instead of establishing virtues that crossed disciplines, I want to demonstrate that Pannekoek's the epistemic virtues in both Marxism and astronomy are connected precisely because they both emerged from a single conception of the self.

Uncovering the relation between Pannekoek's astronomy and socialism is crucial because it has been conspicuously lacking in most historical research conducted on his person. Until now, historians have primarily focused on his political writings, with little—if any—attention to his astronomical research or the relation between the two.[11] This is unfortunate because, as mentioned above, finding common ground between the two might hold the promise of increasing our understanding of both. The lack of interest in the relation between Pannekoek's astronomy and his Marxism can partially be attributed to his own attitude regarding the subject. Outwardly, he tried to keep his two

---

Dear et al., "Objectivity in Historical Perspective," *Metascience* 21 (2012): 11–39; Robert W. Smith, review of *Objectivity*, by Lorraine Daston and Peter Galison, H-Albion, H-Net Reviews (Aug 2012); www.h-net.org/reviews/showrev.php?id=32919 (accessed 7 Apr 2016).

10. Matthew Stanley, *Practical Mystic: Religion, Science, and A. S. Eddington* (Chicago: University of Chicago Press, 2007).

11. Biographical studies on Pannekoek include: Mark Boekelman, "The Development of the Social and Political Thought of Anton Pannekoek, 1873–1960: From Social Democracy to Council Communism" (PhD dissertation, University of Toronto, 1980); Corrado Malandrino, *Scienza e Socialismo: Anton Pannekoek, 1873–1960* (Milan: Franco Angeli Libri, 1987); John Gerber, *Anton Pannekoek and the Socialism of Workers' Self-Emancipation, 1873–1960* (Dordrecht: Kluwer Academic Publishers, 1989); Bart van der Steen, "Anton Pannekoek en het Orthodoxe Marxisme," *Vlaams Marxistisch Tijdschrift* 40 (2006): 73–82. Of these, only Malandrino includes more than a passing mention of Pannekoek's astronomical research.

careers strictly separated. In his astronomical papers, there was never any mention of his political preference, and he hardly ever discussed politics with colleagues.[12] Likewise, in his political writings, his scientific background was rarely mentioned. Evidence of the extent of the separation between the two careers in Pannekoek's own view are his two memoirs that separately discuss his careers in the workers' movement and in astronomy.[13] At the same time, Pannekoek also conceded that some connections did exist between his astronomy and socialism. When, in 1957, fellow council communist Benjamin Sijes inquired about their relation, he answered:

> Interaction existed in so far, that the method of natural science, which I had learned thoroughly, helped me to discover the science of society in Marxism; and that has remained the foundation of my work.[14]

The interaction that Pannekoek alluded to was not an outward, guiding, or causal connection; it was an internal, intellectual one. The prism of epistemic virtues promises to uncover the details of this connection by looking underneath the surface of his work and by focusing on the methodology and epistemology underlying his research.

Pannekoek's idiosyncratic brand of Left Marxism provides another reason why the framework of epistemic virtues is ideally suited to his case. A defining feature of his Marxist philosophy was his elaboration on the role of the human mind in dialectic materialism—an aspect that he felt was lacking in the writings of Marx and Engels. Much of Pannekoek's Marxist writings were aimed at filling this gap by explaining how the human mind processed information and how it turned experience into general abstractions. Because epistemic virtues are often developed as technologies of the self, this aspect of Pannekoek's socialism, which we can justifiably call his philosophy of mind, seems particularly promising. If we uncover what his conception of the human mind was, then we can determine how he constructed his scientific persona to counteract its weaknesses and utilize its strengths.

Of course, the scientific persona is only an ideal type—both a representation of the ideal scientist and a way that scientists want to present themselves to

---

12. Anton Pannekoek, *Herinneringen: Herinneringen aan de Arbeidersbeweging; Sterrekundige Herinneringen* (Amsterdam: Van Gennep, 1982), esp. 238.

13. Ibid. The two memoirs were written in the winter of 1944 and were intended for his family. They were published together with introductions by B. A. Sijes for the socialist memoirs, and Edward P. J. van den Heuvel for the astronomical memoirs.

14. Pannekoek, *Herinneringen* (ref. 12), 16–17.

society—not necessarily an accurate representation of the actual scientist. In the same vein, epistemic virtues prescribed and advocated by scientists do not always concur with their actual scientific practice.[15] It is important, therefore, to focus not only on how Pannekoek envisioned the scientific and scholarly persona in his Marxism, but also on the scientific persona that emerges when we investigate his scientific research. Here again, the case of Pannekoek is especially suited because he spend much of his career studying the Milky Way. Astronomy in general—and Milky Way research in particular—is especially well-suited for a case study in epistemic virtues because it is a predominantly visual science, and as Daston and Galison have shown, the framework of epistemic virtues provides a powerful tool for extracting the underlying epistemological concerns of scientists from the way they visually represent their data.[16]

As with his Marxism, Pannekoek had a peculiar and original approach to galactic astronomy. He did not follow the example of his nineteenth-century predecessors, who attempted to develop all-encompassing models of the distribution of stars through smoothed mathematical formulae. Instead, he acknowledged the irregular and complicated character of the Milky Way and stressed the importance of investigating particularities. Because, so far, very little research has been conducted on Pannekoek's astronomical research,[17] I will provide a detailed analysis of his Milky Way research, not only to extract his scientific persona, but also to determine whether we can situate his contributions within broader developments in contemporary science. Pannekoek's novel methods seem to coincide with the wider shift described by Daston and Galison that occurred throughout all of science during the nineteenth century, away from a focus on the universal that is inherent in truth-to-nature science and toward an appreciation of the particular and idiosyncratic.[18] By looking at the epistemic virtues underlying his astronomical research, we can establish whether this truly the case.

This paper will start with a short description of Pannekoek's life to explore the outward relation between his socialist career and his astronomical career, in an attempt to discover why he chose to keep them separated and see where and how they could have influenced each other. This is followed by a detailed investigation of Pannekoek's Milky Way research. This research can be seen as directed toward two separate but connected goals. For the sake of clarity, these will be discussed separately. First, there will be a discussion of his efforts to provide a complete systematic representation of the visual appearance of the Milky Way. Pannekoek was the first astronomer to provide such a systematic account, and it is especially interesting to explore the multiple complementary techniques he developed toward this goal. Then we will discuss his second goal in galactic research: determining the statistical distribution of stars in the Milky Way system. Here, we will investigate whether we can understand why he took a different approach from not only his predecessors, but also his contemporaries, by reviewing the subject from the perspective of their respective epistemic virtues. Finally, we will explore Pannekoek's Marxist philosophy, especially his theory of the human mind, to investigate whether this will allow us to connect his epistemic virtues in astronomy and Marxism, and provide a more unified description of his entire life.

---

15. Herman Paul, "What is a Scholarly Persona? Ten Theses on Virtues, Skills, and Desires," *History and Theory* 53 (2014): 348–71, esp. 348–54.

16. Daston and Galison, *Objectivity* (ref. 8); Daston and Galison, "Image of Objectivity" (ref. 8).

17. Secondary sources on the astronomical contributions are limited to several eulogies, some entries in biographical dictionaries, and a concise biography; Edward P. J. van den Heuvel, "Antonie Pannekoek (1873–1960): Socialist en Sterrenkundige," in *Een brandpunt van geleerdheid in de hoofdstad*, ed. J. C. H. Blom, P. H. D. Leupen, and P. de Rooy (Hilversum: Verloren, 1992).

18. Daston and Galison, *Objectivity* (ref. 8).

The goal of this paper is, on the one hand, to investigate whether we can connect the socialist philosophy of Pannekoek with his scientific methodology by looking at the epistemic virtues he prescribed and adhered to in both, and see whether this leads to a more unified description of Pannekoek, one that does not view him as an astronomer *or* a Marxist, but as both at the same time. On the other hand, we want to investigate if it is possible, with this unified description, to understand the methodological choices that Pannekoek made in his galactic research, especially in the context of the broader developments in contemporary science.

## A LIFE OF ASTRONOMY AND SOCIALISM

Anton Pannekoek was born on January 2, 1873, in the small rural town of Vaassen in the Netherlands. He born into a middle class family, and as such, he attended the local *Hogere Burgerschool* (HBS) in nearby Apeldoorn.[19] There

---

19. The HBS was a type of secondary education for children of the upper middle class that strongly emphasized the natural sciences. Recently, historians have debated the role played by the HBS in the rise of Dutch science in the late nineteenth century; see, e.g. Bastiaan Willink, "Origins of the Second Golden Age of Dutch Science after 1860: Intended and Unintended he developed his interest in astronomy, and he was persuaded by his teachers to pursue an academic career in the subject. He decided to study in Leiden because of its famous observatory, but was required to take additional courses in the classical languages before being able to enroll, which took him an additional three years to complete. During this period, he started making systematic observations of the night sky, which he documented in personal journals.[20] He started in Leiden in 1891, and graduated five years later. He was employed as a geodesist for three years until he was offered a permanent position as observer at the Leiden Observatory in 1899. While working there, he found the time to work on his dissertation on the light curve of the variable star Algol, which he finished in 1902.[21]

Now that Pannekoek was employed as observer, he followed the expectations of both family and employers by becoming a member of the Liberals in Leiden. Internally, however, he struggled with his political beliefs and the apparent lack of social relevance in his work.[22] He became acquainted with Willem de Graaff, who introduced him to socialist literature. He found himself enthralled by utopian thinkers, which eventually caused him to convert to socialism. It did not take him long to claim his place in the socialist movement. Together with De Graaff, he founded the local chapter of the Sociaal Democratische Arbeiderspartij (SDAP) in Leiden and soon took up the position of chair.[23] Despite his full-time job at the observatory, he was indefatigable in his work for the SDAP; he organized meetings, edited the local weekly paper, and worked long nights in the cooperative bakery, which he had helped finance.[24] Through his party activities, he met his eventual wife Johanna Maria (Anna) Nassau Noordewier, with whom he would have two children. He also started to develop his own socialist philosophy, for which he found an audience through the SDAP's leftist theoretical review, *De Nieuwe Tijd*. In his early articles, we already find many of the topics that would dominate his later work,

---

Consequences of Educational Reform," *Social Studies of Science* 21 (1991): 503–26; Ad Maas, "Civil Scientists: Dutch Scientists between 1750 and 1875," *History of Science* 48 (2010) 75–103.

20. Pannekoek, *Herinneringen* (ref. 12), 229–31; Pannekoek's astronomical journals can be found at API and in the Personal Archive of Anton Pannekoek, Museum Boerhaave, Leiden.

21. Pannekoek, *Herinneringen* (ref. 12), 233–36; Anton Pannekoek, "Untersuchungen über den Lichtwechsel Algols" (PhD dissertation, Leiden University, 1902).

22. He documented these internal struggles in a separate notebook: Anton Pannekoek, Wijsbegeerte en Politiek, Notebook 1898–11 Jun 1899, API.

23. Pannekoek, *Herinneringen* (ref. 12), 71–75.

24. Gerber, *Workers' Self-Emancipation* (ref. 11), 6–11.

in particular the emphasis on the mental factors of materialism.[25] A detailed discussion of these ideas will be provided later in this paper.

It did not take long before a conflict emerged between Pannekoek's new political activism and his professional career as astronomer. In the spring of 1903, he publicly supported and encouraged the protests of railway workers against a proposed law that prevented government personnel from striking. For his role in the failed strike, he was personally reprimanded by conservative Prime Minister Abraham Kuyper, who explained that, as a civil servant, he was not allowed to encourage unlawful behavior.[26] Significantly, Pannekoek's stance in the strike did not go only against government policy but also against that of the SDAP. It was illustrative of his growing criticism of what he perceived to be the opportunistic and revisionist tendencies of the SDAP leadership.[27]

While Pannekoek's international reputation as socialist theorist grew, he was becoming more and more disillusioned with his activities at the Leiden Observatory. Under the directorship of H. G. van de Sande Bakhuyzen, who had a reputation for being resistant to modernization, the once respectable Leiden Observatory had stagnated. The suffocating rigidity was disheartening, and the endless calculations that never seemed to result in publications were enough to frustrate even the always-scrupulous Pannekoek. Eventually a conflict over working hours convinced him to give up his career in astronomy in 1906.[28]

Pannekoek decided to move to Berlin where he had been invited to teach historical materialism at the SPD *Parteischule*, newly founded by Karl Kautsky and August Bebel. Unfortunately for him, this teaching position lasted less than a year before the Prussian government barred him from teaching on the grounds of his non-citizenship. Robbed of his regular income, Pannekoek became a traveling lecturer and created a *Zeitungskorrespondenz*, a weekly newspaper column that was sent to subscribing socialist newspapers.[29] Meanwhile, his interest in astronomy was rekindled as the result of meetings with

---

25. Ibid., 12–21.
26. Pannekoek, *Herinneringen* (ref. 12), 238.
27. Boekelman, "Development" (ref. 11), 45–54; Gerber, *Workers' Self-Emancipation* (ref. 11), 36–42.
28. David Baneke, "Teach and Travel: Leiden Observatory and the Renaissance of Dutch Astronomy in the Interwar Years," *JHA* 41 (2010): 167–98, esp. 169–70; Pannekoek, *Herinneringen* (ref. 12), 234–37.
29. Gerber, *Workers' Self-Emancipation* (ref. 11), 43–46.

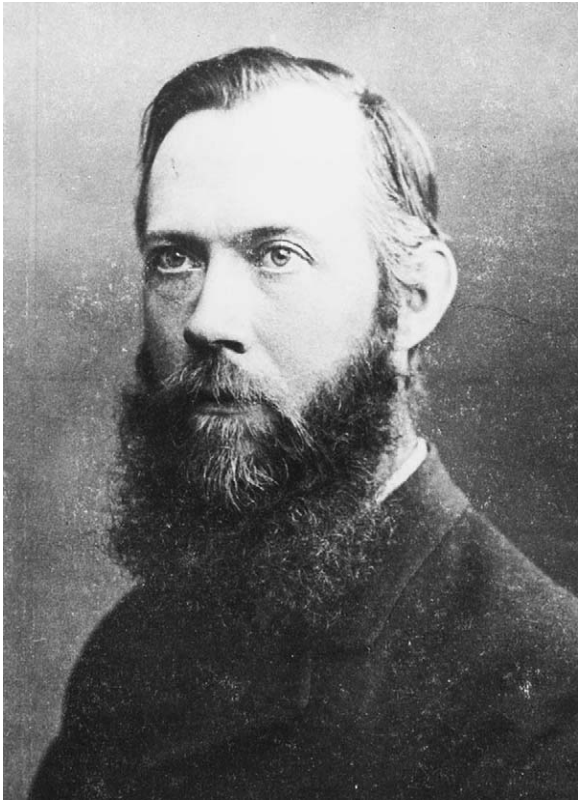

**FIG. 1.** Pannekoek ca. 1908. *Source*: International Institute for Social History, Amsterdam, call nr. IISG BG A10/804.

Karl Schwarzschild and Ejnar Hertzsprung of the Potsdam Observatory. He even managed to publish several astronomical articles while in Germany.[30]

In 1910, Pannekoek moved to Bremen where he was offered another opportunity to teach, this time by the local *Parteischule*. There, he quickly established himself as one of the theoretical leaders of the radical Bremen Left. He criticized both parliamentarism and trade unionism, arguing that the only way a truly democratic socialist society could be formed was through a total destruction of the existing state, initiated by the workers themselves.

---

30. Pannekoek, *Herinneringen* (ref. 12). Schwarzschild had earlier suggested Pannekoek as potential candidate for Professor of Mathematics in Göttingen, a suggestion that was rejected by the commission because Pannekoek was both a foreigner and a socialist; Karl Schwarzschild to Hendrik Antoon Lorentz, 15 Jul 1907, Nachlass Karl Schwarzschild, Briefe 472, Göttingen State and University Library.

He believed that social reforms within the framework of the existing government could never truly liberate the workers but instead merely pacify their natural revolutionary tendencies. His left radicalism eventually led to a controversy with Kautsky, but also gained him the support of Lenin and the Bolsheviks.[31]

When the Great War broke out in 1914, Pannekoek moved back to the Netherlands where he worked as physics teacher for various secondary schools and a part-time lecturer in the history of astronomy at the Leiden Observatory.[32] An opportunity to return to professional astronomy came in 1918, when the director of the Leiden Observatory, E. F. van de Sande Bakhuyzen, passed away unexpectedly. With the support of Kapteyn, Leiden professor of astronomy Willem de Sitter suggested a reorganization of the observatory with himself as director and Pannekoek and Hertzsprung as deputy directors.[33] Although De Sitter and Hertzsprung were hired, the conservative Dutch government prevented the appointment of Pannekoek, arguing that a communist should not be teaching at a state university. De Sitter was not pleased with Pannekoek, but the latter argued that he had fully committed himself to astronomy and that it was not his fault that the government was willing to sacrifice science to politics.[34]

---

31. Gerber, *Workers' Self-Emancipation* (ref. 11), 72–88, 95–100; John Gerber, "From Left Radicalism to Council Communism: Anton Pannekoek and German Revolutionary Marxism," *Journal of Contemporary History* 23 (1988): 169–89, esp. 172–78; H. Schurer, "Anton Pannekoek and the Origins of Leninism," *The Slavonic and East European Review* 41 (1963): 327–44, esp. 329–34.

32. Pannekoek, *Herinneringen* (ref. 12), 240–41.

33. Kapteyn spoke very highly of Pannekoek's astronomical accomplishments but considered him unsuited to run the entire reorganization of the observatory; Wolter Reinold de Sitter, "Kapteyn and de Sitter: A Rare and Special Teacher-Student and Coach-Player Relationship," *Legacy*: 79–108.

34. A detailed account of Pannekoek's rejection in Leiden is provided in: David Baneke, "'Hij kan toch moeilijk de sterren in de war schoppen': De afwijzing van Pannekoek als adjunct-directeur van de Leidse Sterrewacht in 1919," *Gewina* 27 (2004): 1–13. This would not be the last time that the Dutch government interfered: in 1920, Einstein's appointment as special visiting professor in Leiden was delayed because he was mistaken for a communist activist with the same last name, and in 1934, mathematician Dirk Jan Struik was not allowed to become visiting professor at the Delft Institute of Technology because he was a communist; see, respectively, Jeroen van Dongen, "Mirror Images and Mistaken Identity: Albert and Carl Einstein, Leiden and Berlin, Relativity and Revolution," *Physics in Perspective* 14 (2012): 126–77; Gerard Alberts, "On Connecting Socialism and Mathematics: Dirk Struik, Jan Burgers, and Jan Tinbergen," *Historia Mathematica* 21 (1994): 280–305, esp. 281.

The rejection in Leiden proved to be a blessing in disguise for Pannekoek, as he was also offered to found his own astronomical institute at the University of Amsterdam, which was a municipal rather than a state university.[35] Since he could not afford an observatory, Pannekoek decided to follow the lead of Kapteyn in Groningen and dedicated himself to measuring and reducing photographic plates taken by others.[36] He announced to other astronomical institutes that the goal of the Astronomical Institute of the University of Amsterdam, which was formally founded in 1921, was to investigate "the stellar universe and the galactic system."[37] Although this research remained an important topic at the astronomical institute for the next couple of decades, Pannekoek himself soon redirected most of his attention to the newly emerging field of astrophysics of stellar atmospheres. He was appointed associate professor in 1925 and full professor in 1932.

Pannekoek's return to astronomy did not mean he was no longer involved in socialism. He continued writing for Marxist journals and remained in close contact with both the Bremen Left and the Bolsheviks during World War I. He was elated when he heard about the Bolshevik Revolution and the revolutions in Germany. He was even appointed as one of the founders of the short-lived Amsterdam Bureau of the Communist International. At the same time, he was critical: he stressed that the Soviet Union should not be ruled by the Communist Party but by the workers councils, and warned that Bolshevik tactics could not be successful in the far more developed Western European countries. Soon, Lenin had enough of his criticism and strongly denounced Pannekoek and like-minded left-wingers in his now-famous pamphlet *"Left-Wing" Communism: an Infantile Disorder*, singling out the work of Pannekoek as "particularly 'solid' and particularly stupid."[38] For Pannekoek, it was clear that he occupied an increasingly isolated position. He terminated his membership in the Communist Party Holland (CPH) in 1921 and went into a self-imposed six-year-long break. When he resumed writing in 1927, he aligned himself with the new council communist movement that rejected any form of bureaucracy and instead championed complete self-organization by workers themselves. Pannekoek quickly became one of the leading theoreticians of council communism, but the movement would remain small and struggled to gain any support or relevancy.[39] His days of active participation in the workers' movement were over.

---

35. Pannekoek, *Herinneringen* (ref. 12), 246–47.
36. Ibid., 248–49.
37. Anton Pannekoek, Unadressed standardized letter, ca. 1919, Personal Archive of Anton Pannekoek, Museum Boerhaave, Leiden.
38. V. I. Lenin, *'Left-Wing' Communism: an Infantile Disorder* (Detroit: Marxian Educational Society, 1921), on 42. In the pamphlet, he addresses Pannekoek by his pseudonym K. Horner; Gerber, *Workers' Self-Emancipation* (ref. 11), 132–57; Boekelman, "Development" (ref. 11), 259–98.

After his forced retirement during the Second World War, Pannekoek shifted his attention to historical writings. He wrote about the history of astronomy, in which he attempted to link the developments in astronomy with the practical needs and technological advances of the societies in which it was developed,[40] but also on the prehistory of humankind, arguing that primitive humans emerged from animal origins through the interaction of the use of tools, the development of the human brain, and emergence of language.[41] From these historical writings, it is clear that Pannekoek did not pursue the ideal of a pure, objective science independent of external factors. He strongly believed that technical and ideological developments determined the course of scientific research, and he was part of growing group of scientists in the Netherlands who believed that the primary justification of scientific research was to benefit the society.[42]

In light of his historical writings, it might seem odd that Pannekoek kept up the appearance that his astronomy and Marxism were separated from one another. Looking back at his life, however, we can begin to understand why this was the case. At least twice during his astronomical career, he suffered the consequences of the negative perception of his political affiliation. At the same time, his political opponents were quick to cast him aside as a stargazer who had his head in the clouds, or as a rigid mathematician devoid of any human feeling.[43] Separating his careers—if not their methods—was certainly a practical choice, but it is not immediately clear that it was also an ideological one. This provides further justification to the belief that we will be able to find connections between his socialism and his astronomy if we look beneath the surface and focus on his epistemic virtues.

39. Marcel van der Linden, "On Council Communism," *Historical Materialism* 12 (2004): 27–50.

40. See, e.g. Anton Pannekoek, "The Discovery of Neptune," *Centaurus* 3 (1953), 126–37; and Anton Pannekoek, *A History of Astronomy* (London: Allen and Unwin, 1961).

41. Anton Pannekoek, "Antropogenese: Een studie over het ontstaan van de mens," *Verhandelingen der Koninklijke Nederlandsche Akademie van Wetenschappen, Afdeling natuurkunde, Tweede sectie* 42 (1945).

42. David Baneke, *Synthetisch Denken: Natuurwetenschappers over hun rol in een modern maatschappij, 1900–1940*, (Hilversum: Verloren, 2008).

43. Boekelman, "Development" (ref. 11), 16–17.

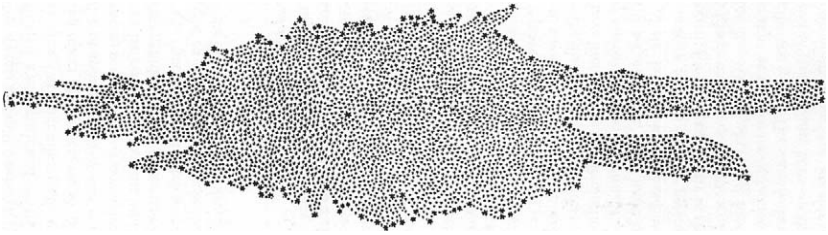

**FIG. 2.** William Herschel's first attempt at a model for the distribution of stars in the stellar system, created in 1785. *Source*: Herschel, "On the Construction of the Heavens," *Philosophical Transactions of the Royal Society of London* 75 (1785).

## EARLY GALACTIC ASTRONOMY

When Pannekoek started his astronomical career, the question of the structure of the stellar system and its relation to the Milky Way was at the forefront of astronomical research. The predominant method of researching this topic was through statistical analysis of the location, apparent magnitude, and proper motion of stars.[44] This research program, known as statistical or sidereal astronomy, can be traced back to William Herschel's 1785 paper "On the Construction of the Heavens," in which he attempted to determine the dimensions of the galactic system by counting stars. He argued that if all visible stars are contained in the galactic system, stars are distributed roughly uniformly, and his telescope could penetrate to the edge of the system, then the number of stars in a certain direction of the sky was a direct indication of the distance to the edge in that direction. The rhombus-shaped system he deduced with this method can be seen in Figure 2. In later life, he came to disavow his three assumptions; his own research on binary stars and star clusters indicated that stars were certainly not uniformly distributed, and his newly constructed 40-foot telescope indicated that the edges of the system were still beyond reach.[45] It was clear that to understand the structure of the galaxy, much more data was needed, and in the following century, many astronomers devoted their attention to counting and measuring the stars.

---

44. Apparent magnitude indicates the brightness of a star as seen from Earth. Lower magnitude indicates that a star is brighter. The brightest stars in the sky are about zeroth magnitude, and sixth magnitude stars are the faintest ones that can be detected by the naked eye. Proper motion is the apparent movement of a star in the sky as seen from the sun.

45. Michael Hoskin, *The Construction of the Heavens: William Herschel's Cosmology* (Cambridge: Cambridge University Press, 2012), esp. 58–74; E. Robert Paul, *The Milky Way Galaxy and Statistical Cosmology, 1890–1924* (Cambridge: Cambridge University Press, 1993), esp. 13–20.

One of the first to make extensive use of this new data was Hugo von Seeliger of the University of Munich. Seeliger had determined that the density of stars as a function of distance from the Sun could be determined by the ratio of the number of stars of a given apparent magnitude and the number of stars one magnitude brighter. In 1898, he calculated this star-ratio for the stars in the *Bonner Durchmusterung*, the most comprehensive star catalogue available for the Northern Hemisphere, and found that it was lower than what would be expected for a uniform distribution of stars, especially in the direction of the galactic poles. In his model, the Milky Way Galaxy was an ellipsoidal system, approximately 10,000 parsecs in diameter along the galactic equator and 1,800 parsecs in the direction of the galactic poles.[46] The sun was placed in the center of this system, and the star density thinned out exponentially toward the edges of the system.[47] Seeliger's method relied heavily on complex mathematical manipulations and theoretical presuppositions. In particular, he required the luminosity function, which describes the relative number of stars as a function of absolute brightness, to be shaped like a Gaussian distribution. Only then was his mathematical analysis valid.

An alternative method for statistical astronomy was developed by Dutch astronomer Jacobus Kapteyn. Unlike Seeliger's mathematical analysis, Kapteyn approached the topic empirically and numerically. One of the first steps taken by Kapteyn was to determine the luminosity function empirically for nearby stars, the first results of which were published in 1902. By assuming that this luminosity was valid throughout the stellar system, he could then use it to compute the density distribution empirically as well. He developed his first model in 1908, when he divided the night sky into three sections: the galactic plane, covering the sky from −20° to 20° galactic latitude; the galactic poles, covering the sky above 40° and below −40° galactic latitude; and the transition zone in between those two sections. For each of these three sections, he determined the star-ratio separately and derived in the star density distribution from it. The results were very similar to those produced by Seeliger: Kapteyn's model too was a flattened ellipsoid with gradually decreasing star density toward the edges.[48] Kapteyn kept refining his results throughout the years, and by the end of his life, his model had become known as the Kapteyn Universe.

---

46. Parsec is a measure of distance, with one parsec being 3.26 lightyears.
47. Paul, *Milky Way Galaxy* (ref. 45), 63–78.
48. Jacobus C. Kapteyn, "On the Number of Stars of Determined Magnitude and Determined Galactic Latitude," *Publications of the Astronomical Laboratory at Groningen* 18 (1908).

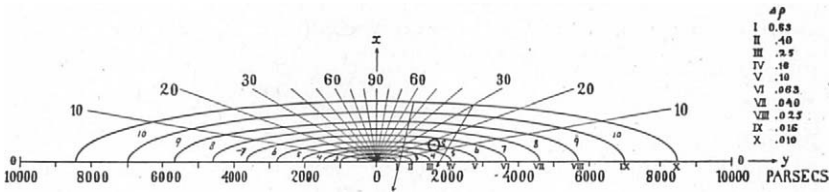

**FIG. 3.** The dynamical model of the galactic system, created by Kapteyn in 1922. The sun is located near the sun and is indicated with a circle. *Source*: Kapteyn, "First Attempt" (ref. 49).

Kapteyn was initially skeptical of the fact that the Sun had such a central place in his system and suspected that this was caused by interstellar extinction. He searched actively for evidence of the existence of interstellar extinction until, in 1916, Harlow Shapley produced results that indicated space was indeed free of interstellar absorption. Safe in the knowledge that his method was indeed valid, Kapteyn eventually came up with a dynamic model of the galaxy, seen in Figure 3, in which the stars had an orbital rotation and the Sun was located very near the center of the system at only 650-parsec distance.[49]

Kapteyn is still fondly remembered by astronomers for his numerical and open-minded approach to statistical astronomy. In the words of Dutch astronomer Adriaan Blaauw:

> Kapteyn's approach was basically different from that of contemporaries such as Hugo von Seeliger and Karl Schwarzschild. The latter proposed certain analytical expressions for the [density and luminosity] functions, as well as for the distribution of observed quantities, and then tried to solve for the parameters involved by means of integral equations. Kapteyn, on the other hand, preferred the purely numerical approach, allowing full freedom for the form of the solution.[50]

Historian of astronomy Elly Dekker summarized:

> [Kapteyn] never sacrificed clarity of treatment or exposure of essential details for elegance of presentation; and, although a mathematician himself by his early training, he strongly disliked treatises in which emphasis lay more on the form of the mathematical expression than on proper evaluation of the basic observation.[51]

49. Jacobus C. Kapteyn, "First Attempt at a Theory of the Arrangement and Motion of the Sidereal System," *ApJ* 55 (1922): 302–28. Paul, *Milky Way Galaxy* (ref. 45), 150–58.

50. *Dictionary of Scientific Biography*, s.v. "Kapteyn, Jacobus Cornelius," by Adriaan Blaauw (1970–90).

51. Elly Dekker, "Jacobus Cornelius Kapteyn (1851–1922)," *Sterrenkijken bekeken: Sterrenkunde aan de Groningse Universiteit vanaf 1614* (Groningen: Universiteitsmuseum Groningen, 1983): 31–42, on 39.

It is undeniably true that Kapteyn took an inductive approach that prioritized the observational data over sophisticated mathematical techniques. This is especially evident in their different approaches to the luminosity function. Where Seeliger postulated an equation that allowed for easy mathematical manipulation, Kapteyn provided a table with observational data to allow the numbers to speak for themselves, in line with the virtue of mechanical objectivity.

At the same time, Kapteyn had his own preconceived ideas about the distribution of stars. Although he insisted that he was building up from below, where others were building from the top down, he still prioritized the shape of the overall system over the existence of individual particularities. He felt confident that these could be ignored because they were only small deviations from the otherwise symmetrical distribution of stars.[52] This decision shows more in common with the ontological concerns of truth-to-nature and, as will be shown, had a crucial impact on the results of his research. It led to criticism from astronomers like Heber D. Curtis, who wrote, "While I am ready to worship Kapteyn's *methods* . . . , I can not, as most astronomers do, fall down and worship all the results which have come out of this mathematical mill."[53] Likewise, Pannekoek also admired Kapteyn's numerical methods, while believing his results were inaccurate, primarily because they did not reflect the visual appearance of the Milky Way:

> [Kapteyn's] results have been obtained by neglecting all differences in surface density of the stars except the mean regular variation with magnitude and galactic latitude. In this regular universe the Milky Way is considered as a continuous belt of feeble light decreasing at both sides, the visual effect of increasing star density with decreasing latitude. An attentive study, however, shows the Milky Way as an extremely irregular series of bright patches and clouds, sometimes divided in two branches, interrupted by dark spaces and connected by long streams.[54]

For this reason, Pannekoek's early research primarily focused on obtaining an accurate, reliable, and complete measurement of the distribution of Milky Way light.

---

52. Owen Gingerich, "Kapteyn, Shapley, and Their Universes," *Legacy*: 191–212, esp. 201.

53. Heber D. Curtis, quoted in, Robert W. Smith, *The Expanding Universe: Astronomy's 'Great Debate,' 1900–1931* (Cambridge: Cambridge University Press, 1982), on 85.

54. Anton Pannekoek, "Researches on the Structure of the Universe: 1. The Local Starsystem Deduced from the Durchmusterung Catalogues," *PUA* 1 (1924) 1–119, on 2.

**REPRESENTING THE MILKY WAY**

The history of accurate visual representations of the Milky Way is considerably shorter than that of statistical astronomy. John Herschel, son of William Herschel, was the first to publish systematic descriptions and drawings of the Milky Way in 1847, based on his four-year trip to South Africa.[55] It took three more decades before similar drawings were published for the Northern Milky Way by Eduard Heis of the University of Münster in 1877, and by Jean-Charles Houzeau, director of the Brussels Observatory, in 1878. Houzeau's atlas, *Uranométrie générale*, was the first to represent the brightness of the Milky Way with isophotic lines, contour lines that represented areas of equal brightness. Another noteworthy drawing of Northern Milky Way was produced by Dutch journalist and amateur astronomer Cornelis Easton in 1893, which was made with assistance from Pannekoek in Leiden.[56] In his introduction, Easton discussed the problems associated with drawing the Milky Way accurately. Because of its extreme faintness, it was not only very difficult to compare the brightness of different parts of the Milky Way, but also deceptively easy to exaggerate the contrast in drawings. Furthermore, he mentioned that, because the Milky Way was comprised of the collective light of many faint stars, it was ultimately a visual phenomenon created by the observer; it could even be called an optical illusion. To make matters worse, the appearance of the Milky Way could be altered easily by observational circumstances or by foreground stars. It was therefore important not to assign too much value to its appearance. Nevertheless, it was still important to keep drawing it, as drawings still provided a valuable opportunity to track changes in the large-scale structure of the Milky Way over time.[57]

Pannekoek himself had been interested in the appearance of the Milky Way from a very young age. His early journals contain multiple observations of features of the Milky Way, which were sometimes accompanied by rudimentary drawings or isophotic maps. This indicates that he already started developing his distinct method of representing the Milky Way before he started his formal education in astronomy. In 1897, when still a student, Pannekoek published a series of articles in several popular astronomical magazines in Germany, England, and the United States, in which he called upon amateur astronomers to make observations of the Milky Way for the purpose of mutual comparison. Observing the Milky Way would have been an ideal way for amateur astronomers, even those without instruments, to contribute to astronomy, especially since a definitive representation of the Milky Way was still lacking. A comparison between the drawings of Easton and those of Irish astronomer Otto Boeddicker with Pannekoek's own revealed that the Milky Way showed different structures in each drawing. Pannekoek was intrigued by this difference of interpretation and speculated on its cause:

> There may be two explanations of these extensive divergences between existing drawings of the Milky Way. They may be the consequence of the different methods employed by the observers, their unequal skill and experience, but it may also be that the character of the galactic phenomenon precludes its being fixed by delineation.[58]

---

55. An assessment of historical descriptions of the Milky Way prior to 1893 is given in Cornelis Easton, *La Voie lactée dans l'hémisphère boréal* (Dordrecht: Blussé, 1893), esp. 11–17.

56. Pannekoek, *Herinneringen* (ref. 12), 234–35.

57. Easton, *Voie lactée* (ref. 55), 1–10.

The drawings of Easton and Boeddicker did reveal "very remarkable agreement"[59] for certain minor parts of the Milky Way, however, so the latter explanation was probably too pessimistic. An accurate representation of the entire structure, upon which they could all agree, should also be possible.

If the differences were the result of the varying skill and methods of the observers, it should be possible to eliminate them. Pannekoek proposed to combine the work of many different independent observers, thereby eliminating the personal quirks of individual observers. This way, it would be possible to create a representation of the Milky Way that was as "true as possible." He recommended a systematic, dual method for observing the Milky Way. The first step was to draw isophotic maps that provided the large-scale brightness distribution of Milky Way light. The downside of these isomaps, however, was that they were poorly suited to capture the precise characteristics of minute particularities. To cover these, the drawings had to be supplemented with detailed verbal descriptions of these particularities.

> [I]ndeed [a verbal description] is much more intelligible to every one than a picture; for a picture gives the opportunity of doubting of what is seen by the observer, especially when the inaccuracies of the multiplying process add to the impossibility of drawing everything exactly as we desire it.[60]

---

58. Anton Pannekoek, "On the Necessity of Further Researches on the Milky Way," *Popular Astronomy* 5 (1897): 395–99, on 397.
59. Ibid., 397.
60. Anton Pannekoek, "On the Best Method of Observing the Milky Way," *Popular Astronomy* 5 (1897): 524–28, on 526.

An important prerequisite for astronomers observing the Milky Way was that they did not acquaint themselves with the earlier descriptions of the Milky Way before making their own. "[The Milky Way's] great faintness makes it very easy to see, what we expect to see: and preconceived ideas will soon vitiate the results."[61] Observers familiar with earlier drawings would inevitably emulate the structure of those drawings, which meant that their drawings would no longer be independent observations.

Pannekoek mentioned two reasons in this paper why representations of the Milky Way were important. The first was that they could be used to track changes in its visual aspect over time. The detailed verbal descriptions were especially well suited for this purpose. The second reason was that the appearance of the Milky Way should be used as a guide and a reference point for sidereal astronomy. On this point, he strongly disagreed with Seeliger and Kapteyn, who both considered individual features less important than the overall symmetry of the statistical distribution of stars.

What we notice in these papers is that, already at a young age, Pannekoek developed a strong opinion on how the Milky Way should be observed. Especially interesting in his method is the role of the observer. To get a representation that was free from the personality and subjectivity of individual observers, multiple representations of different observers had to be combined. Because each observer sees the Milky Way differently, depending on skill, method, and experience, each observer had to observe independently. During their observations, they had to free their minds from preconceived ideas about the structure of the Milky Way because they would inevitably be steered toward the confirmation of these ideas in their observations.

**The Mean Subjective Image**

Pannekoek's call to amateur astronomers had not been successful: only Easton followed the proposed method and supplemented his drawings with isophotic graphs.[62] Pannekoek's own drawings of the Northern Milky Way were not

---

61. Ibid., 524. This was a common concern with astronomers; see, e.g. Albert van Helden, "The Accademia del Cimento and Saturn's Ring," *Physics* 15 (1973): 237–59, esp. 245 and 258; K. Maria D. Lane, *Geographies of Mars: Seeing and Knowing the Red Planet* (Chicago: University of Chicago Press, 2011), esp. 23–63.

62. Cornelis Easton, "La distribution de la lumière galactique compare à la distribution des étoiles catalogues, dans la Voie lactée boréale," *Verhandelingen der Koninklijke Akademie van wetenschappen te Amsterdam (Eerste sectie)* 8, no. 3 (1903). Easton mentioned following Pannekoek's suggestion in an unpublished overview of his work: Cornelis Easton, "Kort overzicht van published until 1920, even though they were made in 1897–99 and 1910–13. Interestingly, he presented his results not in two, but in four different ways. The verbal descriptions and isophotic charts with supplemented by charcoal drawings and numerical values for the surface brightness presented in tables. Most intriguing, however, is that he did not only presented his own results, but also combined them with observations make by earlier astronomers to create what he termed "the mean subjective image" (*durchschnittlich subjectives Bild*). The epistemic concerns regarding these methods were discussed in a separate chapter entitled "Die Milchstrasse als Phänomen."[63]

Pannekoek started by explaining that the Milky Way was an optical phenomenon—the result of the accumulation of light emitted by countless faint stars. The way this optical phenomenon was perceived by the human eye and interpreted by the human mind was influenced by several altering factors.

First, there were the inherent limitations of human anatomy, the so-called optic-anatomical factor. Because of the limited number of nerves in the human eye and the vast amount of stars that form the Milky Way, it was a real possibility that light of multiple stars fell upon a single nerve in the retina. This explained, according to Pannekoek, why the Milky Way was observed as a continuous spotted region of light, rather than the uncountable number of discrete points of light that it actually was.

Another factor was the psychological-physiological factor. The sensibility of the eye and the way its signals are processed by the brain are dependent on the properties of the object that is being observed. When specific patches of the Milky Way are large and bright, it is easier for the brain to detect patterns than when they are small and faint. The patterns detected by the brain were strictly personal according to Pannekoek—every person inherently observed the Milky Way differently from every other person.

Finally, there was the purely psychological factor. Pannekoek claimed that observers were greatly influenced by their own expectations and prior experience. When looking at the same section of the sky twice, it was likely that one would see the same patterns because these were expected to emerge. That same

---

mijn sterrenkundig en meteor. werk," Archief Cornelis Easton (1864–1927), Museum Boerhaave, Leiden, inv. nr. 427b. Contributions by amateur astronomers in the form of drawings was significantly larger in fields like planetary astronomy; see Jennifer Tucker, *Nature Exposed: Photography as Eyewitness in Victorian Science* (Baltimore: Johns Hopkins University Press, 2005), esp. 209.

63. Anton Pannekoek, "Die nördliche Milchstrasse," *Annalen van de Sterrewacht te Leiden* 11, no. 3 (1920), esp. 14–17.

observer could see different patterns, however, if unaware of the fact that it was the same part of the sky. Additional knowledge about a portion of the sky—for example, the existence of a star cloud or a dark nebula—could also alter expectations of the observer, and thus the patterns that could be detected.

More strongly than Easton, Pannekoek stressed that the appearance of the Milky Way was inherently an optical illusion; light streams seemed to appear between rows of stars, patterns seemed to emerge in uniform patches, and once these were seen, they could not be erased. At the same time, this appearance was of crucial importance to understanding the shape and structure of the galaxy. Only if the distribution of the stars matched the appearance of the Milky Way, could it be claimed that they truly formed the Milky Way. How, then, could a reliable, widely acknowledged appearance of the Milky Way be extracted that was valid for all observers?

The main problem, according to Pannekoek, was the purely psychological factor, which not only altered the Milky Way image from observer to observer, but also with the same observer depending on the circumstances of the observation. One way was for observers to attempt to avoid any contact with prior knowledge. In his own observations, Pannekoek purposely avoided consulting even his own earlier observations to avoid biases.[64] Another, as he had suggested in 1897, was to combine the observations of multiple independent observers, for "[t]heir differences give an idea of the objective uncertainties in the faint details, which goes far beyond the limits of subjective security."[65] By averaging all these descriptions, an image could be created that was only altered by anatomical and physiological factors and thus valid for all human beings—the mean subjective image.[66]

Before Pannekoek could start comparing observations, he first had to present his own observations. As mentioned before, these observations were provided in four different forms. There were drawings of the Milky Way that "represent the real aspect of the Milky Way as true as possible"[67] (Figure 4);

---

64. Anton Pannekoek to E. F. van de Sande Bakhuyzen, 26 Feb 1910, Leiden Observatory Archives, directorate E. F. van de Sande Bakhuyzen, Leiden University Library, Leiden, inv. nr. 33: 13–16.

65. Ibid., 16.

66. Interestingly, in earlier communication, Pannekoek used the term "mean objective image" (*gemiddeld objectief beeld*); see Anton Pannekoek to Willem de Sitter, 27 Jun 1919, WdS, inv. nr. 45.1: 56. I was unable to find why he changed from calling it "mean objective" to "mean subjective."

67. Pannekoek, "Nördliche Milchstrasse" (ref. 63), 20.

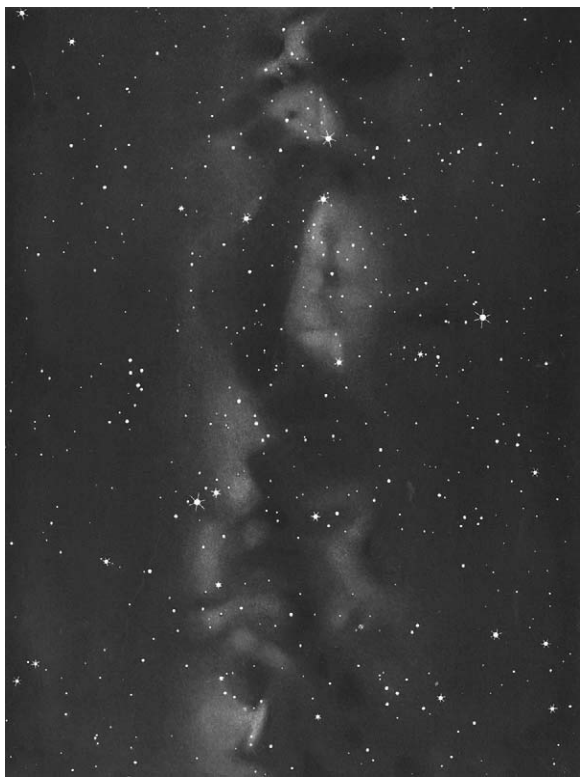

**FIG. 4.** The appearance of a section of the Northern Milky Way drawn by Pannekoek in 1920. *Source*: Pannekoek, "Nördliche Milchstrasse" (ref. 63).

isophotic maps of the same regions (Figure 5); numerical values for the surface brightness; and finally detailed verbal descriptions of individual particularities of the Milky Way. The latter were supplemented with descriptions from earlier observers, including those by Boeddicker, Easton, and Houzeau. In total, these descriptions spanned 72 pages, taking up the majority of the publication.[68]

Combining the drawings of the Milky Way, however, proved to be a challenge. To create the mean subjective image (Figure 6), each drawing had to be quantified so that the arithmetic mean of the different observations could be taken. A major problem was that most of the previous publications were not systematic enough to be quantified easily. Pannekoek decided that only the drawings by Easton and himself could be used in whole. Drawings made by

---

68. Ibid., 18–89.

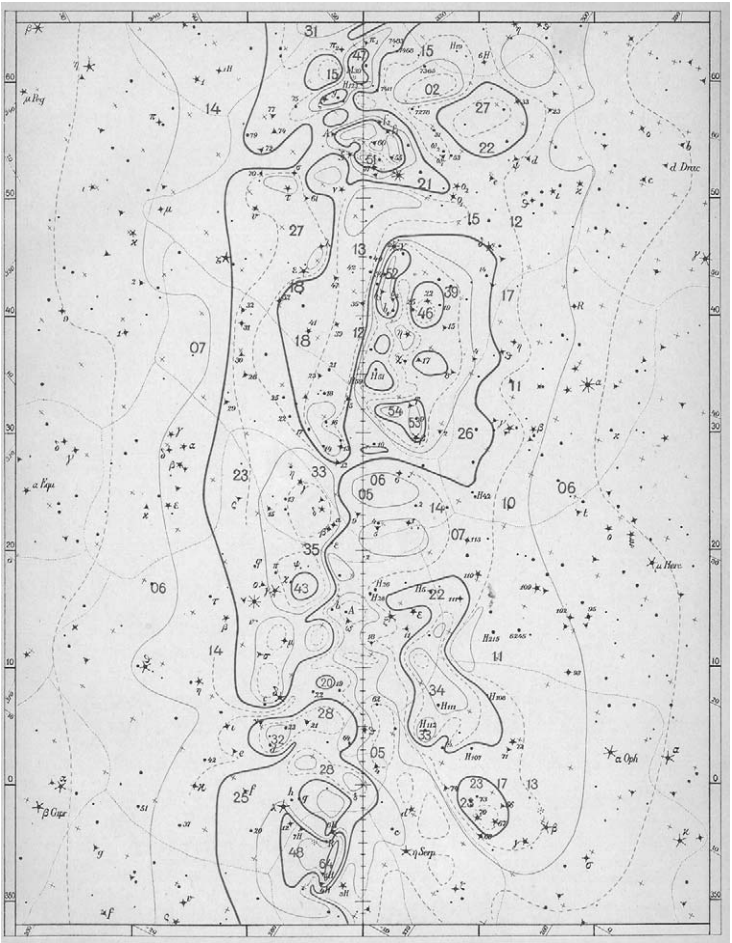

**FIG. 5.** An isophotic map drawn by Pannekoek of the same section of the Northern Milky Was as depicted in Figure 4. The lines indicate areas of equal brightness while the numbers give a numerical value for the brightness at a specific point. *Source:* Pannekoek, "Nördliche Milchstrasse" (ref. 63).

Boeddicker and the German astronomer J. F. Julius Schmidt were useful for clarifying specific feature-rich areas, but not systematic enough for determining the large-scale structure.[69] Pannekoek noticed some interesting differences between the individual drawings of the Milky Way and the mean subjective image. Larger objects often appeared smaller and shallower in the mean subjective image than in the individual images—the result of observers not

69. Ibid., 11–14.

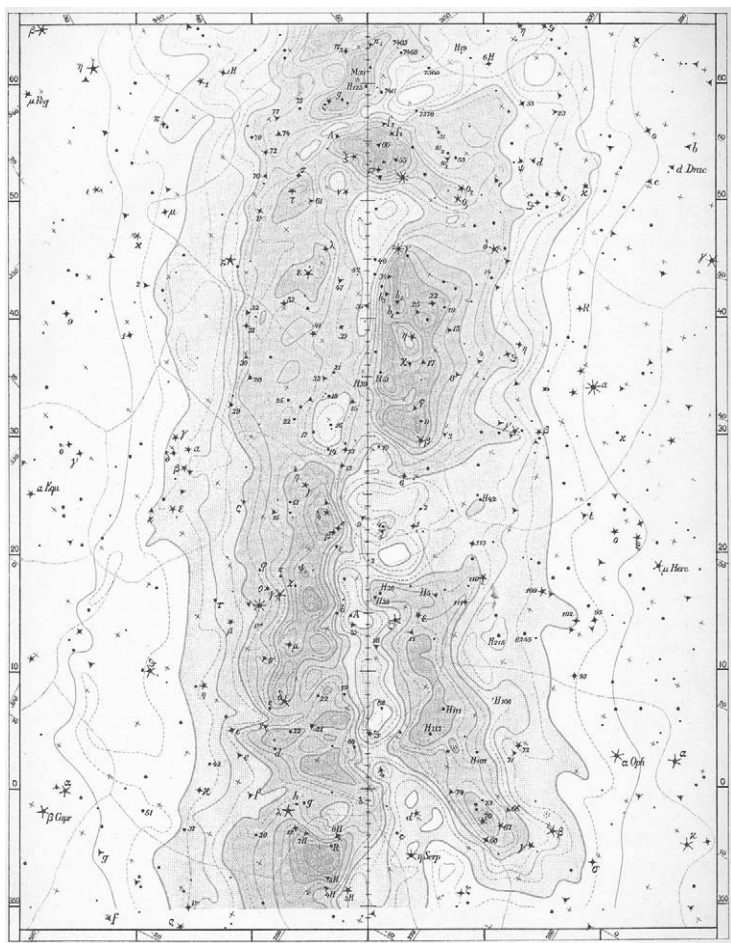

**FIG. 6.** The mean subjective image of the same section of the Milky Way as depicted in Figure 4 and 5. This diagram is created by averaging over several drawings made by independent observers. *Source*: Pannekoek, "Nördliche Milchstrasse" (ref. 63).

agreeing over the exact size and brightness of these objects. For Pannekoek, this was a strong indication that individual observers often exaggerate the sharpness and distinctness of objects.[70]

Pannekoek's development of the mean subjective image of the Milky Way tells us much about the scientific persona he envisioned and the epistemic virtues he advocated. He realized that the aspect of the Milky Way was only an illusion created by the collective light of many faint stars, but, like Easton, he

70. Ibid., 108.

still felt that this illusion had value: it could teach us more about the galactic system and its development over time because it combined the information of many different stars into a single object. As we will see, Pannekoek believed that the goal of natural science was to provide economy of thought from the abstraction of sense perception. Making use of the intuitive way the human eye created the optical illusion known as the Milky Way for practical purposes made perfect sense because it provided exactly this economy of thought. Multiple drawings made by independent observers were needed, however, to make sure that these drawings did not depend too much on personal experience and interpretation. Intuition was a virtue; prior knowledge was not.

There was a striking difference between Pannekoek's epistemic ideas and those of earlier astronomers, like the nineteenth-century nebulae astronomers described by Omar W. Nasim in his book *Observing by Hand*. Those nebulae astronomers believed that multiple observations of the same observer were needed to stabilize the image of a nebula; only through repeated observations could an objective representation be given.[71] One of them, Ebenezer Porter Mason, also made use of isophotic diagrams, but for him, these were not the finished product; they were a way of assisting and guiding the drawing of the brightness distribution in nebulae. For all of them, prior experience and knowledge of individual nebulae played a crucial part in understanding each nebula's idiosyncratic details and avoiding illusion, in stark contrast to Pannekoek's attempts to eliminate the personal touch of the observer.[72] Crucially, there was a difference in the fundamental goal of the drawings. Where Pannekoek wanted to represent the Milky Way as it was perceived by the human eye, the nebulae astronomers aimed for an exact representation of the nebula as it existed in nature. As such, the latter shared their aims and concerns with later photographers of nebulae.[73] Pannekoek, on the other hand, believed that photography could never fully replace human eye observations.

**Extrafocal Photographic Photometry**

The first attempts at taking photographic images of the Milky Way were made in the late nineteenth century, but the results were usually disappointing as the features of the Milky Way resolved into individual stars. Only Edward Barnard succeeded in capturing the Milky Way clouds on photographic plates. He made use of a wide-angle lens, which allowed the light of individual stars to overlap and reproduce the clouds that formed the Milky Way. Although his first photographs were published in 1889, the final work was only published posthumously in 1927 as *A Photographic Atlas of Selected Regions of the Milky Way*.[74]

---

71. Omar W. Nasim, *Observing by Hand* (Chicago: University of Chicago Press, 2013).
72. Ibid., 131–37.
73. Omar W. Nasim, "The 'Landmark' and 'Groundwork' of Stars: John Herschel, Photography, and the Drawing of Nebulae," *Studies in the History and Philosophy of Science* 42 (2011): 67–84.

From these early attempts, consensus grew among astronomers that photographic representations of the Milky Way were fundamentally different from drawings based on naked eye observations; they disagreed, however, about which should be preferred. Edward S. Holden, director of the Lick Observatory, argued that "it seems to be unquestionable that [photography] is the only one which should be employed in the future,"[75] while Easton considered photographs to be superior to drawings, although the latter, especially in the form of isophotic drawings, still retained their value "as an independent evidence, and for certain well-limited purposes."[76] Pannekoek, on the other hand, stressed in 1920 that drawings of the Milky Way could never be replaced by photography because it failed to accurate reflect the surface brightness of the Milky Way light.[77]

Nevertheless, already in 1919, Pannekoek had developed a method that would make a photographic representation of the Milky Way possible. Specifically, this could be achieved through the technique of extrafocal photography. By taking the photographic plates slightly out of focus, the light of a star was spread out over a disk.

> If . . . each starpoint is extended to a circle, the mean surface brightness of the sky over such a circle may be measured by the blackness of the plate; the scale being afforded by the extrafocal images of the bright stars on the plate. Such a picture will bear a much greater resemblance to the visual aspect of the Milky Way than an ordinary photograph.[78]

---

74. William Sheehan, *Immortal Fire Within: The Life and Work of Edward Emerson Barnard* (Cambridge: Cambridge University Press, 2007), esp. 266–77; Pannekoek, *History of Astronomy* (ref. 40), 475.

75. Edward S. Holden, "Considerations on the Methods of Representing the Milky Way, Suggested by a Recent Work," *PASP* 6 (1894): 24–30, on 28.

76. Cornelis Easton, "A Photographic Chart of the Milky Way and the Spiral Theory of the Galactic System," *ApJ* 37 (1913): 105–17.

77. Pannekoek, "Nördliche Milchstrasse" (ref. 63), 15.

78. Anton Pannekoek, "Photographic Photometry and the Colour of the Scutum Cloud," *BAN* 2 (1923): 19–24, on 19.

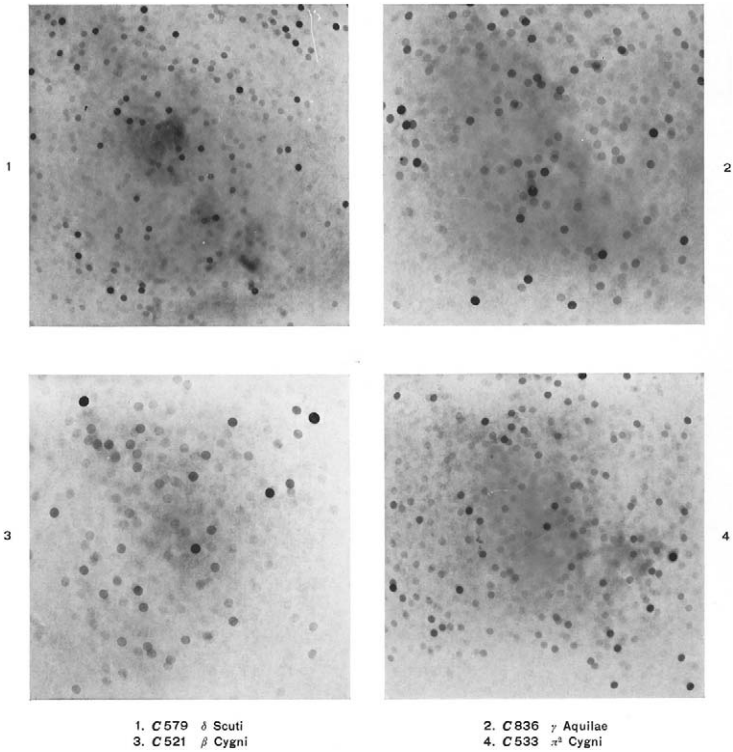

1. C 579  δ Scuti
2. C 836  γ Aquilae
3. C 521  β Cygni
4. C 533  π³ Cygni

**FIG. 7.** Four of the extrafocal photographic plates taken by Max Wolf for Pannekoek to use in his photographic research of the Milky Way. *Source*: Pannekoek, "Photographische Photometrie" (ref. 80).

Pannekoek obtained extrafocal photographic plates from Max Wolf of the Heidelberg-Königstuhl Observatory (Figure 7), and the first results were quite encouraging. Extrafocal photographic plates proved successful in reflecting the brightness distribution of the Milky Way light, and in more detail than the drawn representations could.[79] Unlike Barnard's photographs, Pannekoek's attempt was not meant to convey the appearance of the Milky Way exactly. Instead, its purpose was for photometry: to provide numerical values for the brightness measurement of the Milky Way, which in turn could be used to construct isophotic maps. Pannekoek was so pleased with the method that he later extended it to cover the entire Milky Way, with the northern part being published in 1933 and the southern part in 1949.[80]

79. Ibid., 22.
80. Anton Pannekoek, "Photographische Photometrie der nördlichen Milchstrasse nach Negativen auf Sternwarte Heidelberg (Königstuhl) aufgenommen von Max Wolf," *PUA* 3 (1933);

Pannekoek's extrafocal method reinforces what could already be seen in his drawings—his interest was not in depicting the true structure of the Milky Way but in creating a structure in the distribution of galactic light. For this purpose, the photographic method turned out to be well suited. This did not mean that drawings were no longer valuable, however. Photography was meant to supplement drawings, not replace them.

The fact that Pannekoek still saw value in naked-eye observations of the Milky Way is clear from his effort to extend his research to the Southern Milky Way. An earlier attempt at extending this research had been made by German astronomer Josef Hopmann in 1922, who stated that he had precisely followed Pannekoek's instructions.[81] Pannekoek, however, was critical of the results. He believed that there was a discrepancy between the brightness scales in the two opposite regions of the Southern Milky Way.[82] He was also sceptical of the coarse features of the Milky Way presented in Hopmann's graphs.[83] To extend the research himself, he first came up with the idea to travel to South Africa,[84] but eventually decided to conduct his research in Java when he was invited to be part of the 1926 Dutch solar eclipse expedition to that island.

Pannekoek was stunned with the richness of the Southern Milky Way and realized that the coarse details of Hopmann's maps had indeed been truthful.[85] Not only was the sky different, but also the way he observed it. He described these differences between his observations of the Southern and the Northern Milky Way in detail. For the first time, he was truly able to observe while being completely unfamiliar with the features of the night sky, as he always maintained was the ideal. He quickly found out, however, that this lack of familiarity was not without problems. It took him several nights before he was familiar enough with the southern sky to observe without constantly losing overview and having to re-orientate himself.[86] Furthermore, he realized that he could never completely avoid the effects of prior knowledge, even when

---

Anton Pannekoek and David Koelbloed, "Photographic Photometry of the Southern Milky Way after Negatives Chiefly Taken at the Bosscha Observatory," *PUA* 9 (1949).

81. Josef Hopmann, "Eine Neue Milchstraße," *Astronomische Nachrichten* 219 (1923): 189–200.

82. Anton Pannekoek, "Some Remarks on the Relative Intensities of the Two Sides of the Milky Way," *BAN* 3 (1925): 44–46.

83. Anton Pannekoek to Cornelis Easton, 19 Apr 1926, Archief Cornelis Easton (1864–1927), Museum Boerhaave, Leiden, inv. nr. 427h.

84. Anton Pannekoek to Willem de Sitter, 22 Sep 1923, WdS, inv. nr. 45.2: 61.

85. Pannekoek to Easton, 19 Apr 1926 (ref. 83).

86. Anton Pannekoek, "Die südliche Milchstrasse," *Annalen van de Bosscha Sterrewacht Lembang (Java)* 2, no. 1 (1928), esp. 3–9.

observing a new region of the sky. Increased knowledge of the importance of absorbing nebulae, for example, made observers "more than before inclined to notice and emphasize dark columns and interruptions as real resolved objects."[87] Despite these problems, Pannekoek felt that better knowledge of the method and the way observational data had to be handled resulted in a successful representation of the Southern Milky Way—one that was more certain than his earlier image of the Northern Milky Way.[88]

During his expedition to Java, Pannekoek not only observed, but also prepared for extrafocal photographic plates to be taken at the Bosscha Observatory in Lembang. This further emphasized that he felt naked eye observations and extrafocal photographic images were complementary. In principle, the extrafocal photographic depiction could have been sufficient both for the comparison with the results of sidereal astronomy as well as for tracking changes in the brightness distribution of Milky Way. Yet, Pannekoek saw it not as a replacement of the subjective, interpretive methods; it was a supplement to human interpretation—a fundamentally different representation of the Milky Way phenomenon.

Pannekoek's methods for representing the Milky Way are reminiscent of the epistemic virtue of trained judgement. What we have seen throughout this section is that Pannekoek attributes much importance to the role of the observer in creating an accurate representation of the Milky Way, and despite his emphasis that the Milky Way was only an optical illusion, such a representation was much desired. The ideal observer for this task was free of preconceptions about the structure of the Milky Way, and yet played an active role in the observational process. His human capacity for organizing and systematizing was needed; at the same time his individual subjectivity was to be avoided.

### SIDEREAL ASTRONOMY

One of the main reasons why Pannekoek valued an accurate representation of the distribution of Milky Way light was because it should provide guidance in statistical studies of the structure of the galaxy. He felt that this aspect was being ignored by Seeliger and Kapteyn. His own research on the statistical distribution of stars started after reading Kapteyn's 1908 publication on the distribution of stars.[89] Pannekoek's first paper in this field was a direct reaction to the symmetrical ellipsoidal distribution of stars presented therein. In the introduction, he explicitly mentioned the problem with this model:

> [Kapteyn's] conclusion, however, is *in direct opposition to the appearance of the galaxy*. We see the galaxy as a belt of more or less circular masses, patches and drafts designating a totally different structure. . . . The appearance of the galaxy shows . . . that the zone between +20° and −20° galactic latitude should by no means be treated as one whole. In that way parts of the universe of really great diversity of structure would be mixed up. . . . It may be necessary to take all these different parts together for arriving at an average representation of the distribution of the stars in space, but this is *obscuring* the especially striking *character* of this distribution, which shows in the aggregation of stars into clouds and drifts; and it is giving a false impression of the real Milky Way if the star-density is represented as a simple function of [distance] and [galactic latitude].[90]

---

87. Ibid., 6.
88. Ibid.

Pannekoek's criticism was specifically aimed how Kapteyn had organized the stellar data at his disposal, rather than by his numerical methods. He argued that, by dividing the sky only according to galactic latitude, Kapteyn had already presupposed a symmetry in galactic longitude; the ellipsoidal shape of the resulting system was an artifact of this symmetry. Pannekoek's alternative was to assess the star density distribution separately for particular regions in the Milky Way, as a function of both latitude and longitude. In this way it was possible to determine the star density distribution of individual star clusters, which could then be used to calculate the distance from our Sun to each cluster.

To demonstrate his method, Pannekoek selected five regions to investigate based on the visual aspect of the Milky Way. These were two particularly bright spots in Cygnus and Aquila, two faint parts directly adjacent to these clusters, and a fainter part of the Milky Way as comparison. In the case of the two brighter regions, Pannekoek found that they had significantly more stars than one would expect from Kapteyn's results, especially around ninth and twelfth magnitude. From this, he concluded that it was likely that there were indeed multiple star clusters in the directions of Cygnus and Aquila that caused those brighter regions of the Milky Way. Furthermore, in the case of Cygnus, he found that the higher star density was also present in the adjacent darker part, but only in the case of ninth magnitude stars, not in the case of twelfth magnitude stars. Apparently, there was "no organic relation" between the stars of ninth to eleventh magnitude and the Milky Way clouds, because the cluster of ninth magnitude stars seemed to extend into the dark stroke in Cygnus where the Milky Way phenomenon was absent. Instead, the Milky Way light was probably caused by stars fainter than twelfth magnitude.[91]

---

89. Anton Pannekoek to E. F. van de Sande Bakhuyzen, [early Jun 1910], Leiden Observatory Archives, directorate E. F. van de Sande Bakhuyzen, Leiden University Library, Leiden, inv. nr. 33: 21.

90. Anton Pannekoek, "Researches into the Structure of the Galaxy," *PKAW* 13 (1910): 239–58, on 241–42 (emphasis in original).

Already from this first paper by Pannekoek, it is clear that there are stark differences between his approach and that of Kapteyn. One of the major differences was the perceived goal of sidereal astronomy. For Kapteyn, his research was a first step toward developing a grand scheme that would describe the general distribution of stars in the entire galaxy. In this scheme, the irregularities in the distribution could be discarded because they represented only small deviations from the mean distribution.[92] Pannekoek, on the other hand, emphasized exactly those irregularities and argued that to understand the entire system, we first need to understand how particular areas of the galaxy corresponded with the visual appearance of the Milky Way.[93] Another important difference was the role assigned to the astronomer. For Kapteyn, the astronomer had to minimize his own role in interpreting the data. This could be achieved by using systematically organized sections that eliminated the need for interpretation. Pannekoek, on the contrary—as we have learned repeatedly—constantly emphasized the importance of human judgement in organizing and analyzing the data, although this time, the role of judgement was more implicit than in the case of the Milky Way drawings. Here it meant choosing which areas to investigate, and deciding on the relation between statistical data and the brightness distribution in the galaxy. Where interpretation was a vice for Kapteyn, it was a virtue for Pannekoek.

A major issue with statistical astronomy was the lack of complete and homogenous data. Published star catalogues often registered only the position of stars, not their apparent magnitudes. That meant the actual number of stars of a certain magnitude had to be calculated by determining a limiting magnitude for each catalogue: the magnitude of the faintest stars still included in the catalogue. Even then, the limiting magnitudes were far from systematic because the star counts did not always cover the entire area homogeneously. This was a significant problem because the statistical method of Pannekoek and Kapteyn relied heavily on knowing the relative number of stars for various magnitudes. As a solution to this problem, Pannekoek proposed a photographic method for obtaining star counts. Multiple wide-angle photographs would have to be taken of a single region, using geometrically increasing exposure times. By increasing the exposure times geometrically, the limiting magnitude would increase by a constant number with each photograph. The exposures could be taken on a single photographic plate with the plate being slightly shifted in between exposures. This way, the magnitude for each star could easily be determined by the number of times it appeared on the plate.[94]

---

91. Ibid., 256–58.
92. He justifies discarding the deviations in the star distribution in Kapteyn, "Number of Stars," (ref. 48), 2–3.
93. Pannekoek explicated this difference in approach in his memoirs: Pannekoek, *Herinneringen* (ref. 12), 243 and 247.

In 1910 and 1911, Pannekoek received photographic plates of the Aquila region taken by Hertzsprung. In processing these plates, Pannekoek decided to divide the area into 100 squares, which were grouped into five regions. As can be seen in Figure 8, these regions did not have regular shapes. Instead, it seems that the squares were grouped according to a combination of star density and location, with Sections I and II having the most stars, while Section V was relatively poor in stars.[95] Although there were large differences in the total number of stars, Pannekoek found no significant difference in the star-ratio. Denser sections had more stars at every magnitude, rather than at a few magnitudes, as would be expected in the case of star clusters. According to Pannekoek, this indicated the number of distant stars was actually consistent for the entire area. The relatively low number of stars in Section V was probably caused by a triangle-shaped dark nebula, rather than an actual deficiency of stars. This nebula had to be located close enough to darken all but the brightest stars. Rather than forming an "organic" connection with the distant clouds of the Milky Way, the nebula was "only accidently projected" in front of it.[96] Despite these results, Pannekoek felt that the photographic method was inadequate. It proved impossible to penetrate much further than

94. Anton Pannekoek, "A Photographic Method of Research into the Structure of the Galaxy," *PKAW* 14 (1912): 579–84. The Dutch version of the paper was published in 1911.

95. Anton Pannekoek, "Investigation of a Galactic Cloud in Aquila," *PKAW* 21 (1919): 1323–37, esp. 1332.

96. Ibid., 1334. Pannekoek would later made specific attempts at measuring the size and shape of dark nebulae according to their effects on star counts; see, e.g. Anton Pannekoek, "The Distance of the Dark Nebulae in Taurus," *PKAW* 23 (1921): 707–19.

TABLE I. Number of stars.

| I | | | | | | | | | |
|---|---|---|---|---|---|---|---|---|---|
| 121 35 18 1 1 0 | 98 36 15 5 2 0 | 123 44 19 8 5 1 | 97 27 17 7 5 1 | 83 24 11 5 1 0 | 113 44 17 7 0 0 | 100 33 18 7 2 0 | 112 46 27 11 3 1 | 127 51 22 13 6 2 | 118 34 15 9 1 0 |
| 135 40 17 7 1 0 | 88 34 14 3 1 0 | 64 28 17 8 3 0 | 96 38 13 6 4 0 | 101 36 13 5 2 2 | 96 28 13 5 1 0 | 84 24 12 8 5 2 | 107 46 14 6 4 2 | 85 31 20 5 2 0 | 94 37 13 2 1 0 |
| 83 27 14 6 2 0 | 96 33 15 6 1 1 | 97 27 12 5 4 0 | 123 36 20 7 3 0 | 115 40 16 8 3 2 | 91 28 15 6 4 1 | 90 34 20 11 5 1 | 90 35 21 7 5 2 | 77 26 13 7 3 1 | 111 44 21 8 1 0 |
| III 76 25 13 4 0 0 | 76 23 12 4 1 0 | 102 49 25 8 5 2 | 105 32 14 6 2 1 | 77 35 19 8 3 1 | 87 49 19 8 3 1 | 95 35 15 6 0 0 | 120 35 14 6 1 0 | 144 47 16 5 0 0 | 117 53 25 13 4 1 II |
| 52 27 18 8 3 0 | 87 30 12 4 2 0 | 76 23 9 3 0 0 | 94 44 19 8 2 1 | 59 30 18 10 4 3 | 69 30 13 7 1 0 | 92 34 21 8 6 1 | 103 40 18 8 4 1 | 97 46 21 12 3 1 | 127 39 18 9 5 0 |
| 52 18 9 3 1 0 | 78 25 8 4 2 0 | 92 IV 31 9 3 2 0 | 72 22 10 5 1 0 | 84 33 19 6 1 0 | 82 39 20 7 2 1 | 111 32 17 7 3 2 | 110 33 21 5 2 0 | 84 34 16 6 2 0 | 118 40 11 3 0 0 |
| 55 13 8 3 2 0 | 73 21 11 2 1 0 | 88 30 10 4 1 0 | 55 16 4 1 0 0 | 58 21 13 7 3 0 | 99 35 16 9 5 2 | 95 49 25 10 4 1 | 117 41 15 5 2 0 | 106 55 24 9 4 2 | 88 36 17 8 2 1 |
| 45 20 9 6 2 1 | 52 25 10 2 1 0 | 54 25 12 5 2 0 | 19 12 6 0 0 0 | 28 10 5 0 0 0 | 72 24 12 6 3 0 | 89 29 15 5 3 0 | 88 34 20 6 2 0 | 87 32 10 2 1 0 | 94 27 12 3 2 1 |
| 53 16 9 3 3 2 | 51 10 6 3 2 2 | 38 10 6 5 1 0 | 38 9 5 2 1 0 | 21 5 2 0 0 0 | 29 9 5 1 1 0 | 84 26 14 8 2 2 | 63 35 16 9 5 0 | 61 33 15 7 4 1 | 71 30 14 6 5 2 IV |
| V 39 19 12 1 0 0 | 24 13 8 2 1 0 | 35 8 6 4 1 0 | 53 15 8 2 0 0 | 34 11 6 3 2 1 | 39 15 10 4 2 1 | 39 18 8 1 1 0 | 60 30 15 8 2 0 | 46 20 14 5 2 1 | 61 31 17 9 3 0 |

**FIG. 8.** This diagram indicates the star counts of the photographic plates taken of Aquila. The top number shows the number of starts visible at 1900 seconds exposure time, and the following numbers indicate the number of stars that were also visible with exposure times of 600, 190, 60, 19, and 6 seconds, respectively. The bold lines represent the division of the area into five equally large sections. *Source*: Pannekoek, "Galactic Cloud in Aquila" (ref. 95).

the fourteenth magnitude, as increasing noise levels made plates with even longer exposure times unreliable.[97]

It is interesting to notice that, even in a strictly systematic photographic scheme, Pannekoek felt the need to intervene with the organization of the

97. Pannekoek, "Galactic Cloud in Aquila" (ref. 95), 1337.

stellar counts. He organized the field into intuitively determined sections following interesting features—dense clouds in section I and II, and a dark void in section V—and decided that they should be investigated separately. This division allowed him to compare the different sections, which in turn led him to postulate the existence of a dark cloud in the region. This need to interfere with the organization of data can easily be understood in light of his epistemic virtues. If the data were simply organized mechanically, valuable information might be lost that the human mind could intuitively grasp and use to create structure.

### The Distance to the Milky Way

To make better use of the limited data available for stars fainter than fourteenth magnitude, Pannekoek created a model of a single star cluster placed in an otherwise uniform galaxy. For this model, he could compute a theoretical star count, which could be fitted to observed star counts by adjusting the distance and size of the theoretical cluster. This method had the advantage that a fairly precise measure of the distance could be provided with only a limited amount of data; the downside, however, was that small variations in the measured counts could have a significant effect on the final results. In 1919, he used this method to derive distances of 40,000 parsecs to the cluster that formed the Cygnus cloud, and 60,000 parsecs to the cluster that formed the Aquila stream.[98] This result was especially significant because it firmly placed these branches of the Milky Way beyond the limits of both Seeliger's and Kapteyn's systems, which were both less than 20,000 parsec in diameter. Despite using the same basic numerical techniques, Pannekoek had now found results that directly contradicted the model of Kapteyn.

Pannekoek was not the only one to challenge the models of Seeliger and Kapteyn. A year earlier, in 1918, American astronomer Harlow Shapley had published the results of his investigation on the distribution of globular clusters. Shapley had discovered that these clusters seemed to form a system that was distributed symmetrically around the galactic plane, from which he drew the conclusion that they outlined the extent of the entire galactic system. The most striking result, however, was that the center of this system seemed to be at a distance of 20,000 parsecs from the Sun, with the entire system stretching

---

98. Anton Pannekoek, "The Distance to the Milky Way," *MNRAS* 79 (1919): 500–07, esp. 504.

some 100,000 parsecs.[99] Pannekoek explicitly recognized that their results were complementary:

> The results we have arrived at here are in accordance with [Shapley's results], as they place some of the bright parts of the Milky Way at a distance of 40–60,000 parsec. So the starry masses of the galaxy are spread over space as far as the remotest clusters, and clearly both belong together to one system. In this system the dense agglomerations of stars are spread over a flat disc of about 2000 parsec thickness, and in the empty space above and below it the globular clusters are dispersed."[100]

The agreement between their distance scales was not the only reason why Pannekoek believed that Shapley's system could be correct. He also suggested that Shapley's result reflected the appearance of the Milky Way because the eccentric position of the sun explained why the Milky Way in Sagittarius—where the center of Shapley's system was located—was much brighter than in the opposite direction toward Perseus.[101]

The expanded galaxy of Shapley was a much-debated topic in the years immediately after its publication. Although Seeliger rejected Shapley's results in private correspondence, he did not actively participate in this debate.[102] Instead, most of the criticism on the expanded galaxy came from Dutch astronomers who—except for Pannekoek—initially remained loyal to Kapteyn's smaller model. One of Kapteyn's students, Willem Schouten, provided his own measurements of galactic clusters conducted with traditional statistical means and found much smaller distances than Shapley, while Kapteyn and Pieter van Rhijn directly challenged the Shapley's use of Cepheids as a measure of distance.[103] Kapteyn also argued that Shapley's eccentric position of our solar system was difficult to accept because of the symmetry in the star density distribution in all directions of galactic latitude. He felt that, where Shapley was building from above, Kapteyn was building from below.[104] The task of challenging the distances that Pannekoek derived for the Cygnus and Aquila clouds was left to none other than Easton.[105]

---

99. Harlow Shapley, "Globular Clusters and the Structure of the Galactic System," *PASP* 30 (1918): 42–54, esp. 48–50.

100. Pannekoek, "Distance" (ref. 98), 507.

101. Ibid.

102. Helge Kragh, *Masters of the Universe: Conversations with Cosmologists of the Past* (Oxford: Oxford University Press, 2015), on 49 and 60 n. 7.

103. E. Robert Paul, "The Death of a Research Programme: Kapteyn and the Dutch Astronomical Community," *JHA* 12 (1981): 77–94.

104. Gingerich, "Kapteyn, Shapley" (ref. 52).

Easton was no stranger to the two areas under investigation. Already in 1895, with the cooperation of Pannekoek, he had compared the brightness distribution of the Milky Way light in his own drawings from 1893 with the distribution of stars in the Bonner Durchmusterung. He determined that the intensity of galactic light corresponded well with the distribution of stars, and concluded that most likely a real connection existed between the distribution of faint and bright stars.[106] This result was the basis of Easton's criticism of Pannekoek's distance measurements. In his reaction to Pannekoek, published in 1921, Easton stated that "it is obviously improbable that a real condensation of stars in the neighbourhood of the Sun and an extremely distant galactic cloud should be seen almost exactly in the same direction without their being physically related."[107] It was much more likely that the brightest stars of the galactic clouds revealed themselves already at ninth magnitude, while the bulk presented itself only around twelfth magnitude. As evidence, Easton presented a comparison between star counts in the brighter sections and fainter sections of the Milky Way. Three aspects stood out in this comparison: in dense sections there were more bright stars; there was a higher proportion of intrinsically bright B and A stars; and the average proper motion was lower. These results all pointed toward the conclusion that distant galactic stars revealed their presence among the brighter stars. Subsequently, the fact that, in the case of Cygnus, the effect was already noticeable at the ninth magnitude, indicates that these galactic clouds must then be much closer than Pannekoek had calculated.[108]

Pannekoek felt Easton's arguments were inconclusive at best. His main counterargument was that the correlation between the distribution of bright stars and galactic light was not as strong as Easton had presented. To illustrate his point, Pannekoek created a diagram (Figure 9) that compared the distribution of galactic light with star counts from the Bonner Durchmusterung, representing the bright stars, and John Herschel's star gauges, representing the faint stars. He argued that the diagram clearly showed that the distribution of the galactic light corresponded well with Herschel's star gauges but not with the number of *Bonner Durchmusterung* stars. He felt reinforced in his theory that there were two clusters—one nearby cluster within a few hundred parsecs of the Sun, which revealed itself in the *Bonner Durchmusterung* stars, and one distant galactic cloud, which could be seen as galactic light and in John Herschel's star gauges. One concession he was willing to make was that he had overestimated the distance to the galactic cloud, which he now calculated to be only 18,000 parsecs.[109]

---

105. Unfortunately, it is unknown how Kapteyn reacted to Pannekoek's system. He never published a reaction, and both had their correspondence destroyed during the Second World War.

106. Cornelis Easton, "On the Distribution of the Stars and the Distance of the Milky Way in Aquila and Cygnus," ApJ 1 (1895): 216–21.

107. Cornelis Easton, "On the Distance of the Galactic Star-Clouds," MNRAS 81 (1921): 215–26, on 224.

108. Ibid., 222–26.

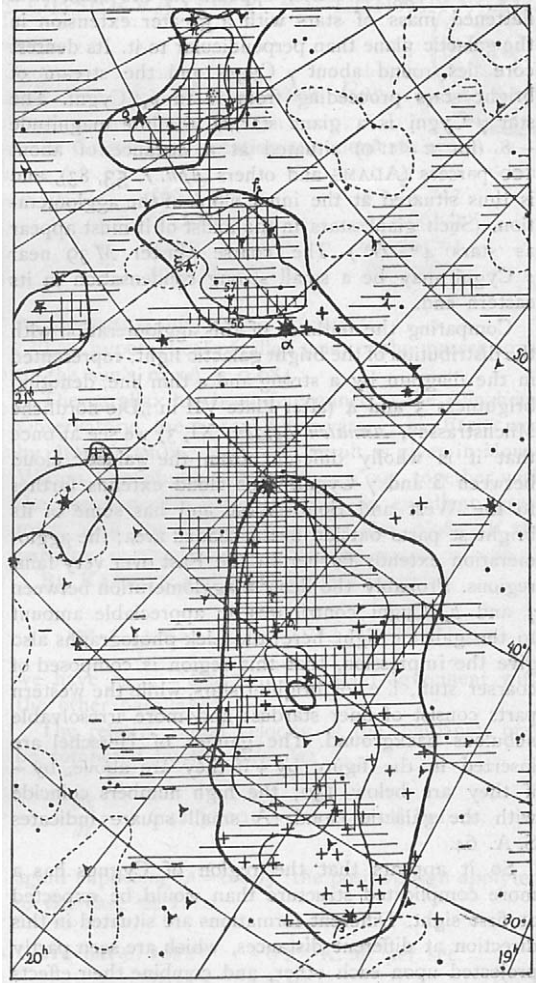

**FIG. 9.** Pannekoek's diagram of the Milky Way in Cygnus. The distribution of Milky Way lines is indicated by isophotic lines, and the distribution of stars is indicated by shaded areas and dotted lines. Denser shading indicates a higher number of stars, and the shaded areas indicate fewer stars. Pluses and minuses indicate whether the star gauges by John Herschel revealed a high or a low number of stars. *Source*: Pannekoek, "Distance Galaxy in Cygnus" (ref. 109).

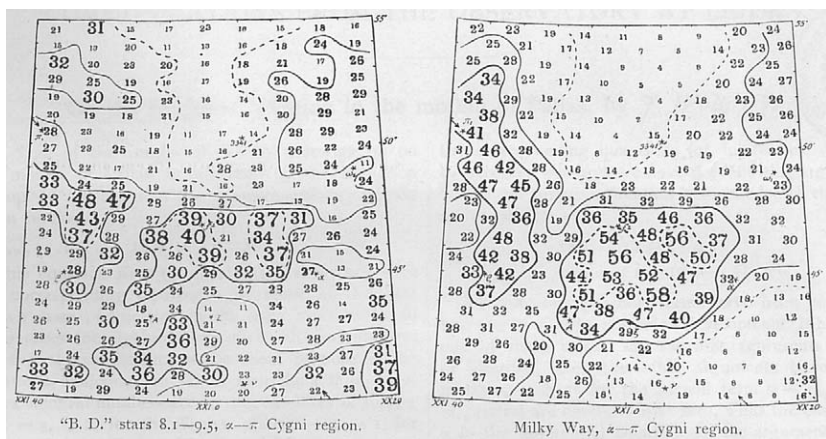

**FIG. 10.** Easton's diagrams of the Milky Way in Cygnus. The left diagram indicates the number of stars in the *Bonner Durchmusterung*, and the right diagram gives the numerical value of the brightness of the Milky Way. *Source*: Easton, "Correlation" (ref. 110).

Pannekoek's diagram did not persuade Easton, however, as the latter made clear in a final paper on the topic in 1922. Easton repeated his earlier argument that the correlation between *Bonner Durchmusterung* stars and galactic light distribution was too strong to be coincidental and presented his own chart as illustration (Figure 10). In the discussion of the charts, Easton appealed strongly to the common sense of the reader:

> We cannot of course expect a perfect agreement, but who could believe that these two diagrams represent two distinct and independent agglomerations of stars, situated respectively at distances of 400 and 18,000 parsec?[110]

Interestingly, Easton published his paper in *BAN*, which was normally reserved for professional astronomers working at one of the astronomical institutes in the Netherlands. Easton persuaded Pannekoek, however, that he should be allowed to publish his paper in the same journal as the paper that had attacked him.[111]

---

109. Anton Pannekoek, "The Distance of the Galaxy in Cygnus," *BAN* 1 (1922): 54–56.
110. Cornelis Easton, "Correlation of the Distribution of Bright Stars and Galactic Light in Cygnus," *BAN* 1 (1922): 157–59, on 159.

The debate between Pannekoek and Easton illustrates how epistemic virtues can influence observers in their research. Easton urged astronomers to use their common sense and to avoid being led astray by misinterpretations of minor deviations. He repeatedly emphasized how "remarkable" a "chance coincidence"[112] of two superimposed clusters would be, and with his diagrams, he hoped that the reader would recognize the rough similarities between the two distributions. In contrast, Pannekoek appealed to the active judgement of the observer while warning against predetermined ideas. His diagram allowed a more direct comparison of the data, and he left it to the reader's own judgement to reflect on the differences in the distributions and determine that they are in fact not that similar. Implicit in this debate were the different roles of the astronomer. Where Easton expected the astronomer to use theoretical training and common sense, in line with the epistemic virtue of truth-to-nature, Pannekoek expected the astronomer to be a consciously intervening, yet unbiased judge of empirical data.

**Switch to the Local System**

Shapley's extended galaxy was also challenged from the United States, most prominently by Curtis of the Lick Observatory. Curtis had done photographic research on novae in spiral nebulae and found that these novae, on average, appeared to be a hundred times more distant than novae located in the Milky Way. That would mean that these novae—and as a result the spiral nebulae—were located well outside the borders of the galaxy. Curtis concluded that these spiral nebulae were "island universes," independent star systems that were similar in size to our own galaxy.[113] At first, the existence of island universes seemed at odds with Shapley's extended galaxy. Supporters of the latter believed that it was large enough to incorporate the entire universe including the spiral nebulae, whereas those who believed in the island universe theory

---

111. Anton Pannekoek to Willem de Sitter, [late May 1922], WdS, inv. nr. 45.2: 21–22.
112. Easton, "Correlation" (ref. 110), 157.
113. Michael A. Hoskin, "Richie, Curtis and the Discovery of Novae in Spiral Nebulae," *JHA* 7 (1976): 47–53; Robert W. Smith, "Beyond the Galaxy: The Development of Extragalactic Astronomy 1885–1965, Part 1" *JHA* 39 (2008), 91–114, esp. 109.

considering spiral nebulae to be much smaller than the size Shapley calculated for the Milky Way Galaxy.[114]

What both theories had in common was the reduced status of the Kapteyn Universe: it was either only one of many independent spiral nebulae, or it was only a small part of a vastly bigger system. Statistical astronomy had proven incapable of providing a cosmological model of the entire universe. Robert Paul claimed that this fundamental failure marked the end of statistical astronomy as a major research program.[115] Although galactic structure remained one of the primary research topics among Dutch astronomers, most no longer followed the statistical methods of Kapteyn.[116] An exception to this was Pannekoek.

The main reason why Pannekoek was able to continue with Kapteyn's methods was that he never had the goal to derive a single model for the *entire* universe. His interest was always in how the collection of particularities, such as star clusters and dark nebulae, together formed the Milky Way phenomenon. He emphasized this difference in approach in his biographical memoirs:

> I strongly sympathized with the work of Kapteyn, but always felt I viewed it differently. He treated the star density distribution only as a function of distance and galactic latitude, ignoring the variance in galactic longitude. Throughout my youth, I always watched the Milky Way and never perceived it as a gradually thinning ellipsoid, but rather as [a collection of] individual accumulations, similar and equally important as Kapteyn's local system. I saw it as my duty to follow my own ideas and determine the structure of the Milky Way as a collection of corporal clouds and streams.[117]

When statistical astronomy was understood to be fundamentally incapable of providing a cosmology, Pannekoek was not deterred, because it could still provide insight into the structure of the galaxy, even if only on a local level. When he published his first results in this reconfigured field in 1921, the title "The Local Starsystem" was indicative of the change of perspective in sidereal astronomy.[118]

---

114. For detailed analysis of the Great Debate, see: Michael A. Hoskin, "The 'Great Debate': What Really Happened," *JHA* 7 (1976): 169–82; Smith, *Expanding Universe* (ref. 53), 77–90.

115. Paul, "Death of Research Programme" (ref. 103), 89.

116. Woodruff T. Sullivan III, "Kapteyn's Influence on the Style and Content of Twentieth Century Astronomy," in *Legacy*: 229–64. For the development of Dutch astronomy in the twentieth century, see David Baneke, *De Ontdekkers van de Hemel: De Nederlandse sterrenkunde in de twintigste eeuw*, (Amsterdam: Prometheus Bart Bakker, 2015).

117. Pannekoek, *Herinneringen* (ref. 12), 247.

118. Anton Pannekoek, "The Local Starsystem," *PKAW* 24 (1922): 56–63. The Dutch version was published in 1921.

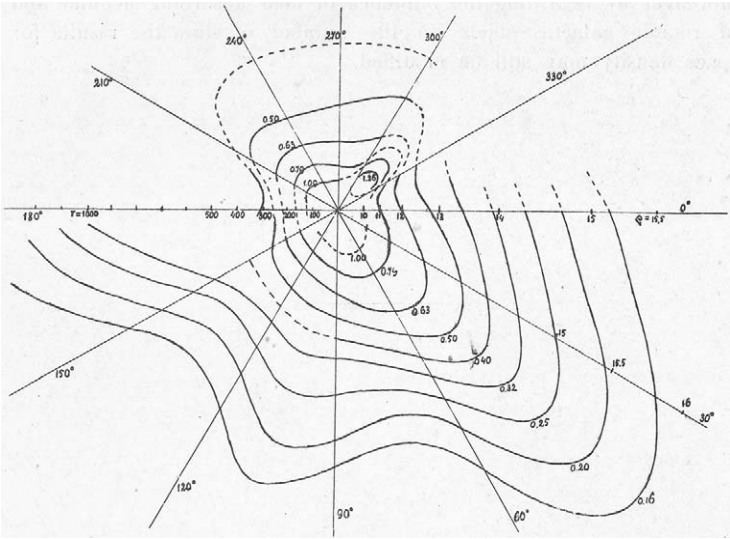

**FIG. 11.** A top-down projection of the star density distribution in the galactic plane. The curved lines indicate areas of equal brightness. *Source*: Pannekoek, "Local Starsystem" (ref. 118).

As in his earlier statistical research, Pannekoek relied on the calculation of star-ratios to determine the distances of star clusters. A marked difference in approach, however, was that he was not looking at specific clusters. Instead, he investigated the density distribution of the entire galactic zone between −20° and 20° galactic latitude as a function of both galactic longitude and distance. As we can see in Figure 11, he tried to accomplish this by dividing the galactic plane into twelve sections of 30° longitude each. The results, which extended to about 1,000 parsecs from the Sun, were presented through lines of equal star counts on a cross section of the galactic zone. Notable features in the distribution were the condensations that appear in the direction of Cygnus, but especially those in the direction of Scorpio (around 315°–330°), where the density rose to 1.25 times the average density around the sun at a distance of 100–200 parsecs.[119] This research was later extended by Dutch astronomer Egbert Albert Kreiken, who managed to cover far larger distances, especially in the Southern Hemisphere.[120]

It might seem as if, with this new method, Pannekoek was moving away from his initial plan to investigate specific irregularities of the Milky Way.

---

119. Ibid., 83–85.
120. E. A. Kreiken, "The Density-Function in the Milky Way," *MNRAS* 85 (1925): 499–507.

After all, by dividing the sky into only twelve parts, he condensed large portions of the sky in each section. He argued, however, that this is not the case because he still determined the star count for each section separately, instead of trying to smooth the distribution. His approach still focused on deviations from the mean and the particular structure of the density distribution—aspects that would remain hidden in an approach that smoothed out the data.

Moreover, this research was only the first step to providing a "third approximation" that determined the distribution of stars as a function of latitude, longitude, and distance. By doing so, it was possible to "[treat] the different galactic features as special objects and [determine] the distribution functions separately for all these special regions in the sky."[121] This new research was not limited to the galactic equator alone, but covered the entire sky. The term "third approximation" may seem to indicate that Pannekoek was simply adding more detail to Kapteyn's "second approximation" of providing the density function as a function of galactic latitude and distance alone. This, however, would neglect the significant differences in the fundamental goal of their research. Pannekoek was not looking for a single star density distribution that could capture the shape of the entire system;[122] he was looking for specific features and irregularities that indicated the existence of star clusters in the solar neighborhood. Whether these clusters formed a complete system or were only a small part of a larger system was not of immediate concern. The acquired knowledge would be useful either way. This attitude, as we have seen, can be explained by his particular epistemic virtues that shunned the desire for grand schemes and instead called for the creation of a structure that could summarize experience and make predictions for the future. In light of these goals, there was no reason for Pannekoek to abandon his research, despite the rapidly changing theories of the universe, and even when other astronomers saw no benefit in continuing this line of research.

In practice, the third approximation meant that Pannekoek calculated the density distribution of stars in each direction of the sky by investigating the relative star-ratios for three separate visual magnitudes, 5.7, 7.4, and 8.6. The results were represented as azimuthal projections of the Northern and Southern Hemispheres for each magnitude (Figure 12). Parts of the sky where the density distribution deviated strongly from the mean were highlighted in the diagram: red for relatively dense areas and blue for relatively sparse areas. The primary conclusion of this investigation was that no central condensation of stars could be found that acted as the center of the system. Instead, he found a number of accumulations and clusters that were roughly comparable with each other in size and density. such as those in the directions of Cygnus, Monoceros, and Carina. For each cluster, the size and density were determined.[123] The results were placed in a schematic top-down projection of the stars in the solar neighborhood that can be seen in Figure 13.

---

121. Pannekoek, "Researches 1" (ref. 54), 2.

122. The original publication in 1924 only covered the sky down to –65° declination. The final southern part was added in 1929 in: Anton Pannekoek, "Researches on the Structure of the Universe: 3. The Cape Photographic Durchmusterung," *PUA* 2, no. 2 (1929): 71–87.

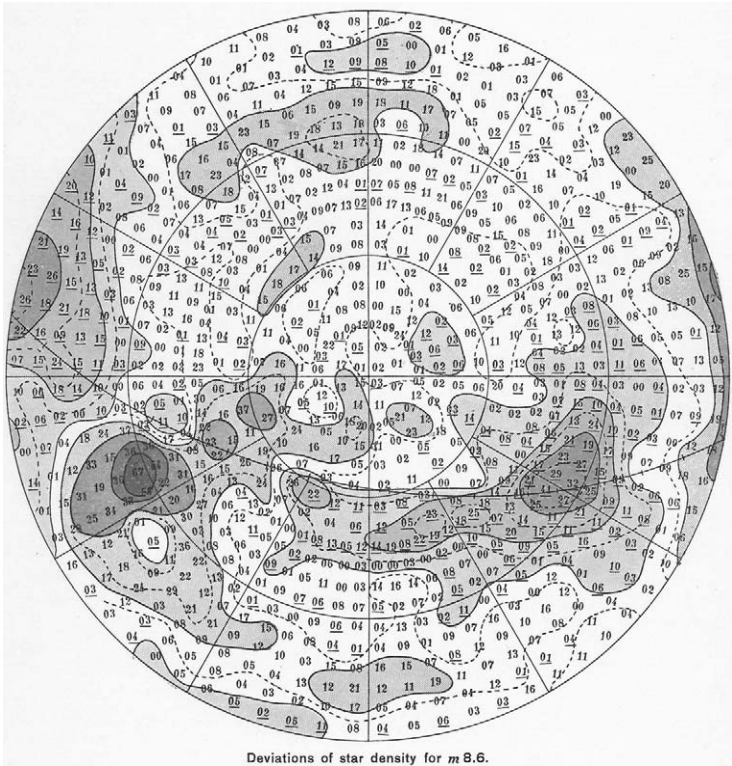

**FIG. 12.** The density distribution of stars of visual magnitude 8.6 in the Northern Hemisphere. Areas of relatively high density were indicated by increasingly darker shades of red, while fainter areas were indicated with blue shading. *Source*: Pannekoek, "Researches 1" (ref. 54).

123. Pannekoek, "Researches 1" (ref. 54).

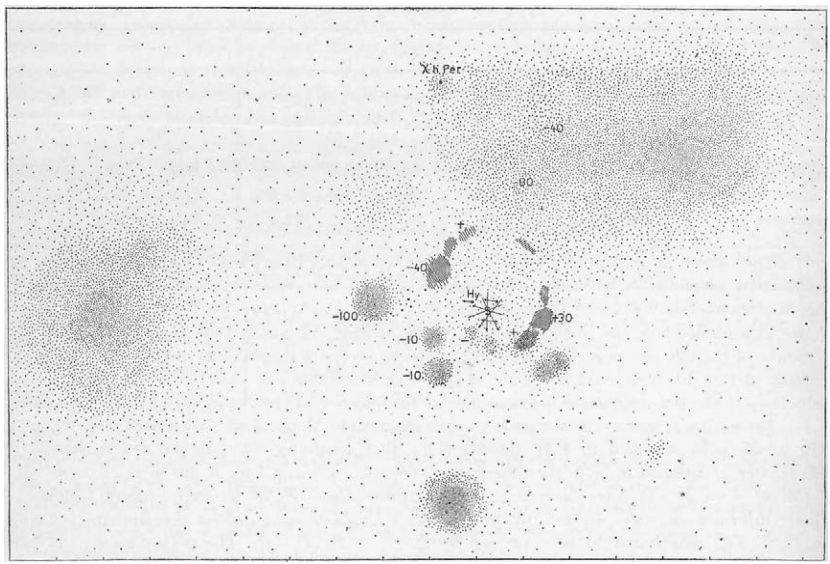

**FIG. 13.** A top-down projection of the distribution of stars in the local system. The numbers on the diagram indicate how far clusters are located above or below the galactic plane. *Source*: Pannekoek, "Researches 1" (ref. 54).

In the preceding pages, we have seen how Pannekoek managed to adapt his research to the changing opinions on the structure of the universe. His adaptability in this matter can be directly linked to the ontology that emerged from his epistemic virtues. His focus was on recreating the distribution of stars from the bottom up, as an accumulation of smaller particular features, and not on formulating a top-down model that could provide insight in the size and shape of the entire system. When faced with the choice, he decided to stay true to Kapteyn's statistical method of deriving star densities from star counts, rather than to Kapteyn's goal of determining the size and shape of the entire galaxy, like the other Dutch astronomers did. For Pannekoek, the realization that statistical astronomy could provide a model for only a small portion of the universe did not diminish its value since it provided a way to understand our particular local cluster in great detail. Furthermore, this knowledge would also have implications on the understanding of the entire Milky Way Galaxy.

Pannekoek continued conducting statistical astronomy until the late 1920s. When Jan Oort found evidence that the Milky Way Galaxy was a large, differentially rotating spiral galaxy, Pannekoek immediately attempted to find the central masses that could be responsible for the rotation.[124] He also attempted to add additional detail to the structure of the local system by searching for clusters of specific classes of stars.[125] In 1930, however, Robert Trumpler published a paper in which he provided definitive evidence for the existence of interstellar absorption.[126] The interstellar matter responsible for this absorption turned out to be irregularly distributed throughout the entire system, meaning that it was nearly impossible to correct for its effect in statistical studies, even for such a limited scope as the local system. Whereas the various larger systems of Shapley, Curtis, and Oort provided interpretive difficulties that were easily integrated within Pannekoek's methodology, the discovery of interstellar absorption provided a practical objection to sidereal astronomy to the point where it could no longer be defended. The features of the Milky Way that Pannekoek had been chasing turned out to be shadows.[127]

## PANNEKOEK'S SOCIALIST THEORY OF KNOWLEDGE

While Pannekoek was exploring new methods to investigate the Milky Way Galaxy, he was also exploring methods to investigate the development of human society. As with galactic astronomy, Pannekoek's most active period in Marxism was during the first few decades of the twentieth century; and as with his representations of the Milky Way, he formulated the philosophical foundations of his work early in this period. Even though his conception of the socialist revolution changed over time, these foundations remained largely consistent. In this section, we will take a closer look at Pannekoek's specific brand of Marxism and investigate how they led to his left radical interpretation of communism, before analyzing how it relates to his epistemic virtues in astronomy.[128]

---

124. Anton Pannekoek, "On the Possible Existence of Large Attracting Masses in the Centre of the Galactic System," *BAN* 4 no. 125 (1927): 39–40.

125. Pannekoek, "Researches on the Structure of the Universe: 2. The Space Distribution of Stars of Classes A, K and B," *PUA* 2, no 1 (1929): 1–70.

126. Robert J. Trumpler, "Absorption of Light in the Galactic System," *PASP* 42 (1930): 214–27; Daniel Seeley and Richard Berendzen, "The Development of Research in Interstellar Absorption, c.1900–1930," *JHA* 3 (1972): 52–64, 75–86, esp. 81–84.

127. Pannekoek, *Herinneringen* (ref. 12), 247.

128. An analysis of Pannekoek's Marxist persona and how he interacted with other Marxist is provided in: Chaokang Tai and Jeroen van Dongen, "Personae and the Practice of Science: Anton Pannekoek's Epistemic Virtues in Astronomy and Socialism," *BMGN—Low Countries Historical Review* 131 (2016): 55–70.

Pannekoek's Marxist philosophy was largely based on the work of two philosophers, Karl Marx and German self-taught philosopher Joseph Dietzgen. Marx' main contribution, according to Pannekoek, was discovering the scientific method of researching society—the materialist conception of history. In fact, he used the terms "Marxism," "social science," and "historical materialism" interchangeably. The basic assumption underlying this method was that people's thoughts and actions were fully determined by social conditions, which meant that to study society was to investigate how material conditions determined societal developments.[129] At the same time, Pannekoek felt that Marx had failed to explain *how* exactly these thoughts were determined by material causes.[130] He believed that this problem was solved by Dietzgen, who had managed to raise philosophy to the level of natural science, not by speculating on the nature of the human mind, but instead by treating the human brain as a naturally evolved organ with a specific function: to make sense of sense perceptions.[131]

Pannekoek conceived the external world as a continuous and infinitely varied collection of ever-changing phenomena, which humans could perceive through their senses. The human mind was tasked with bringing order into the overwhelming stream of sense impressions by abstracting and generalizing this information into distinct categories and concrete concepts. Pannekoek explained it as follows:

> "The mind is the faculty of generalization. It forms out of concrete realities, which are a continuous and unbounded stream in perpetual motion, abstract conceptions that are essentially rigid, bounded, stable, and unchangeable.... The world is a unity of the infinitely numerous multitude of phenomena and comprises within itself all contradictions, makes them relative and equalizes them. Within its circle there are no absolute opposites. The mind merely constructs them, because it has not only the faculty of generalization but also of distinguishing.[132]

Everything that entered the human mind was instinctively transformed into mental concepts and concrete objects; the process was not only influenced by direct sense perceptions, but also by our prior experience in the world. Because it worked instinctively, however, the mind was not always aware of all the material factors that influenced this process.[133] This was especially true because, according to Pannekoek, material factors also encompassed mental factors, including ideas, such as tradition and religion, and social structures, such as social classes and the educational system:

> [The surrounding real] world is not restricted to physical matter only, but comprises everything that is objectively observable. The thoughts and ideas of our fellow men, which we observe by means of their conversation or by our reading are included in this real world. Although fanciful objects of these thoughts such as angels, spirits or an Absolute Idea do not belong to it, the belief in such ideas is a real phenomenon, and may have a notable influence on historical events.[134]

---

129. Anton Pannekoek, "De filosofie van Kant en het Marxisme," *De Nieuwe Tijd* 6 (1901): 549–64, 605–20, 669–88, esp 613.

130. Anton Pannekoek, "Dietzgenismus und Marxismus," *Bremer Bürgerzeitung*, 12 Nov 1910.

131. Anton Pannekoek, "Dietzgens Werk," *Die Neue Zeit* 31 (1913): 37–47.

132. Anton Pannekoek, "The Position and Significance of J. Dietzgen's Philosophical Works," introduction to Joseph Dietzgen, *The Positive Outcome of Philosophy*, translated by Ernest Untermann (Chicago: Charles H. Kerr & Company, 1906): 7–37, on 33.

These mental factors were of crucial importance to Pannekoek. Only by recognizing their influence was it possible to research and comprehend society through scientific methods. Societal developments were caused by the interactions across the traditional categories of mind and matter. Thoughts and ideas were influenced by the material world, but they also acted upon their environment in return.[135] The task of social science was to explain both how ideas emerged as the result of economic, social, and ideological conditions, and how they subsequently influenced these conditions.[136]

One of the most powerful forces in society, according to Pannekoek, was "social memory," which meant "the perpetuation of collective ideas, systematized in the form of prevailing beliefs and ideologies, and transferred to future generations in oral communications, in books, in literature, in art and in education."[137] He claimed that social memory was being used by the bourgeoisie to indoctrinate the working class into endorsing bourgeois ideals and supporting bourgeois interests. The only way that the working class could shake off this indoctrination was through education in historical materialism and activism in the form of spontaneous and self-organized strikes and

133. Anton Pannekoek, "Socialism and Religion," *International Socialist Review* 7 (1907): 546–56, esp. 551–52; Pannekoek, "Filosofie van Kant" (ref. 129), 670.

134. Anton Pannekoek, "Society and Mind in Marxian Philosophy," *Science and Society* 4 (1937): 445–53, on 451.

135. Ibid.

136. For Pannekoek "social science" was synonymous with "Marxism" and "historical materialism," and he used the terms interchangeably.

137. Pannekoek, "Society and Mind" (ref. 134), 453.

demonstrations. Activism would reawaken the revolutionary spirit of the workers, educate them about how they should be organized, weaken the structure of the existing class-based society, and give them self-confidence in their ability to create a new, truly democratic society.[138]

By the mid-1910s, Pannekoek had completely rejected parliamentarianism and trade unionism as paths toward the new society. He believed that any top-down organization would inevitably turn its leadership into a labor aristocracy, who would benefit from perpetuating the status quo within the existing government.[139] Instead, the workers should organize themselves from the bottom up, by congregating into factory councils that together formed the new council communist government. On this point, we can find an analogy with his astronomical research. His bottom-up conception of the ideal society is reminiscent of the bottom-up method he applied in sidereal astronomy, where individual stars congregated into clusters and the combination of clusters formed the Milky Way Galaxy. In both cases, there was no need for an overarching, top-down system to control the basic structure.

More importantly, with Pannekoek's Marxist philosophy in mind, we gain a better understanding of the epistemic choices he made in his astronomical research. The human mind was inextricably involved in abstracting knowledge from the overwhelming variety of sense perceptions. To avoid bias and personal subjectivism, however, it was important to be aware of potential presuppositions and make active use of one's own active judgement. By understanding of the role of the mind, it was possible to maximize its greatest strength—its ability to systematize and organize information—while avoiding its weakness—its susceptibility to bias from prior experience. In representing the Milky Way, this meant being aware of the fact that it was inherently an optical phenomenon created by the light of innumerable stars, and actively searching for structure in the distribution of light. In sidereal astronomy, it meant taking individual particularities seriously and constantly comparing statistical results with the appearance of the Milky Way. Astronomers should embrace the systematizing tendencies of the human mind for their own benefit. This is especially true because, even if it were somehow possible to prevent the human mind from interfering with the registration and collection of phenomena, then it would still only lead to a multitude of particularities without structure or order. Such an unsorted collection of events would have no value in helping to understand the world. Not only was human intervention inevitable, it was desirable.

---

138. Gerber, *Workers' Self-Emancipation* (ref. 11), 95–100; Gerber, "From Left Radicalism" (ref. 31), 174–78.

139. Schurer, "Origins of Leninism" (ref. 31), 333–35.

Pannekoek's active methodology was not limited to Marxism and astronomy either. It was part of his entire philosophy of science. The problem with mechanical materialism, which he associated with the bourgeoisie, was that it attributed a real existence to natural laws of science. This, he argued, was a gross misinterpretation of its actual nature. Natural laws were nothing more than refined experience, created by summarizing sense information and abstracting it into general expressions that indicate what we expect to happen in the future. They did not refer to the properties of real objects, but only addressed the properties of universal, abstract entities. By identifying concrete and singular facts with universals, they were crafted into laws. Both universal entities and natural laws existed only as abstractions in human thought; they were fabricated constructions that brought order to the overwhelming stream of sense perceptions.[140] Like contemporaries such as Ernst Mach and Pierre Duhem, Pannekoek asserted that the goal of science was simply to provide economy of thought by capturing experience and expectations as efficiently as possible. Natural laws should not be interpreted as pre-existing laws waiting to be discovered, but as crafted artifacts that summarize what has happened or what we expect to happen.[141] With that perspective, not only the Milky Way was an optical illusion; any scientific theory was essentially an illusion created by the human mind to allow it to comprehend the world. As long as a theory served a practical goal as an analytical tool, it was worthy of investigation.

## CONCLUSIONS

We started by asking whether an investigation of the epistemic virtues and scholarly personae of Pannekoek would lead to a deeper understanding of his Milky Way research in the context of contemporary developments in science, and in relation to his Marxist philosophy. To start with the latter, the biographical section made it clear that Pannekoek kept his two careers outwardly separated for practical reasons—not necessarily ideological ones. His political stance worked against him at various times during his astronomical career, and his astronomical background was sometimes used by opponents in their polemics against him during his political career. The remainder of the paper, however, revealed where those connections *can* be found. There are striking resemblances between the epistemic virtues that Pannekoek advocated in both fields.

---

140. Anton Pannekoek, "Twee natuuronderzoekers in maatschappelijk-geestelijke strijd," *De Nieuwe Tijd* 22 (1917): 300–14, 375–92, esp. 382; Anton Pannekoek, "Das Wesen des Naturgezetses," *Erkenntnis* 3 (1933): 389–400.

141. Pannekoek, "Twee natuuronderzoekers" (ref. 140); Pannekoek, "Dietzgen's werk" (ref. 131).

Throughout this paper, we have seen that both the astronomical persona and the Marxist persona, as conceptualized by Pannekoek, were actively involved in systematizing and analyzing the infinitely varied stream of information from the external material world. By abstracting this information into natural or social laws, they could summarize what was already known and make predictions about the future. How this worked in practice depended on the field of research the persona was engaged in. The galactic astronomer had to look for the structure in the Milky Way and determine how this coincided with the clustering in the distribution of stars. The Marxist theorist had to analyze how material factors—social, economic, or ideological—determined the behavior of social classes, and develop tactics that would optimize the odds for creating a truly democratic society. In both cases, it was imperative that these personae were open-minded; they should not let themselves be guided by preconceived ideas about how the structures should look, or how the revolution should play out, because such preconceptions would alter their perceptions and would lead them to see what they expected to see. This suggests that, rather than there being some fundamental disconnect, Pannekoek's conception of the ideal scientist and the ideal Marxist were both rooted on the same epistemic concerns, each adapted to suit its own field of research. This, of course, should hardly be surprising because they were ultimately created by the same person: Pannekoek.

Pannekoek's personae can be understood in light of his ontology and philosophy of mind, which he most thoroughly explained in his socialist writings. According to this ontology, the external world was a continuous and infinitely varied stream of events. The human mind could never fully access or understand this stream, and accordingly, this should not be the goal of science or socialism. Instead, he argued that the focus of both should be on how this stream of information was ordered, systematized, and understood by the human mind. The structure created by the human mind made it possible to analyze nature and obtain economy of thought or predict the future. From Pannekoek's writings, we can deduce that for him, the difference between natural science and socialism was primarily determined by the subject matter:

whereas natural science covered only physical matter, socialism also had to take into account ideological, social, and economical relations. This difference explained why meaningful results in the social sciences could only be obtained through a dialectical philosophy, whereas for natural science they could also be obtained through a mechanical philosophy: the processes deviated from each other only when the human mind itself was part of the research subject. Pannekoek's ontology provides us with a helpful guide for understanding his idiosyncratic methodology in astronomy. His idea that science should look for structure in the sense perceptions explains why he developed various methods of representing the Milky Way. These various representations were different ways in which the distribution of stars could be structured; they represented independent research objects that highlighted different aspects of this distribution. The fact that these structures did not exist outside the human mind was irrelevant if they could help us understand the phenomenon of the Milky Way.

Although this study revealed strong relations between Pannekoek's epistemic virtues in his approach to socialism and science, they are certainly not limited to these virtues alone. It is possible, for example, to draw a tentative—but certainly noteworthy—analogy between his model of the local system and his model for council communism. In both cases, he rejected the top-down model that emphasized the overarching system as a meaningful entity. Instead, he utilized a bottom-up method that emphasized the way in which individual persons or stars congregate into larger systems. The collection of these individual clusters provides sufficient structure for the system as a whole without requiring an additional overarching layer.

Finally, there is the strong belief in his methodology even when it became controversial. Here, however, there is also a stark difference. In Marxism, he never seriously questioned the Marx-Dietzgen synthesis he created within a few years after the labor movement. His ideas evolved—at least early in his career—according to the changing social conditions, but they were always founded upon his idiosyncratic interpretation of Marxism. In his statistical astronomy research, on the other hand, he did continue his statistical research of star counts even after it was revealed that it could not be used to determine the structure of the entire galactic system, but when strong practical objections showed that the method was simply not viable, he had no trouble giving it up.

It is important to emphasize that we did not find evidence of any sort of causal connection for Pannekoek's approach to socialism and astronomy. His investigations into socialist philosophy were not decisive for his ontology; in his earliest writings—from when he was still nominally a liberal—we can already find traces of the focus on mentally created structure over externally existing entities. Nor does it make much sense to say that his particular branch of Marxism was evoked by his scientific training; the majority of his close allies lacked this training, and many of his fellow Marxist scientists had markedly different interpretations of socialism. However, finding a causal link was never the goal. What we wanted to discover was whether we can come to a better understanding of Pannekoek's contributions toward astronomy and socialism through viewing him as a complete, unified person, and the examples above indicate that this is the case.

Investigating the epistemic virtues of Pannekoek also greatly helped with the second goal of this paper, which was to situate his astronomical research within the greater developments of contemporary science. Here, Daston and Galison's grand narrative provides useful context for Pannekoek's novel methods. His rejection of the methods of Kapteyn and Easton can be understood in light of typical rejections against truth-to-nature. He accused Kapteyn of sorting the star counts according to his own preconceived ideas. By doing so, Kapteyn could find no other shape for the system than the one he already had in mind. On the other hand, Pannekoek had strong ontological and epistemological objections to mechanical objectivity. The ontological objection was that the external world was an infinite stream of infinitely varied information, which did not make sense without human analysis, so letting nature speak for itself would provide us with no meaningful knowledge at all. The epistemological objection was that by idealizing a fully mechanical approach to science, the most defining aspect of human beings—their ability to analyze—would be lost. Instead, Pannekoek can be seen as one of the earliest adepts of the twentieth century epistemic virtue of trained judgement. He was part of a growing movement of scientists who increasingly emphasized the need for interpreted structure and systematized data. His ideal astronomer was actively involved in systematizing and analyzing the information provided by instruments or sense perceptions. His task was to recognize characteristic or distinguishing features of particularities and highlight them for other astronomers. This strongly reminds us of his socialist philosophy of mind. The very nature of the human mind is to organize and distinguish sense perceptions; this ability to make sense of the infinitely varied external world intuitively is its greatest strength. If scientists are to understand the world and provide economy of thought, they ought to use this strength.

Yet—obviously—upon closer inspection there are also ways in which our story reveals the weaknesses of Daston and Galison's narrative. Kapteyn is difficult to place as he seems to occupy a position that floats somewhere in between truth-to-nature and mechanical objectivity, and Pannekoek is hardly the archetypical trained judgement expert, since he assigns no specific importance to the role of professional training in developing scientific intuition. Of course, this is always the case with grand historical schemes: when they are applied to specific case studies, the strict categories of these schemes fall apart in a sea of context, and it turns out to be impossible to capture the overwhelming complexity within the history of science in a single big picture. This does not mean, however, that these grand schemes are without value. Despite its flaws, the storyline of epistemic virtues, as described by Daston and Galison, helped to situate the development of Pannekoek's astronomy within the broader historical context. It helped to highlight the differences between him and Kapteyn, which led them to contradicting results despite their use of the same mathematical methods and statistical data. It helped to understand why Pannekoek continued to contribute to a field of research that was essentially declared dead several years earlier.

Reflecting on method, I believe that the framework of epistemic virtues provided a valuable perspective for investigating Pannekoek. It showed its strength most clearly in allowing us to find the connections between his astronomical and Marxist research. It helped us to discover the common ground in their epistemology and in their personae. Most of all—and hopefully further biographical research will make this even clearer—it helped to see beyond the separation between disciplinary boundaries that Pannekoek himself created between his astronomical and socialist careers, and opened the door to a more unified and complete description of his entire professional life—a description that recognizes that behind Pannekoek the astronomer and Pannekoek the Marxist there is, in fact, a single person with a consistent conception of the self.

### ACKNOWLEDGEMENTS

This research has been partially funded by Stichting Pieter Zeeman Fonds. I would like to thank Jeroen van Dongen, Edward van den Heuvel, Alexei Kojevnikov, Robert, W. Smith, Léjon Saarloos, H. Floris Cohen, Herman Paul, Ralph Wijers, and two anonymous reviewers for their valuable advice and insightful comments on earlier drafts of this paper.